# An efficient and robust all-Mach consistent numerical scheme for simulation of compressible multi-component fluids including surface tension, cavitation, turbulence modeling and interface sharpening on compact stencils

Yu Jiao[*,1], Steffen J. Schmidt[1], Nikolaus A. Adams[1,2]

[1] Chair of Aerodynamics and Fluid Mechanics, School of Engineering and Design, Department Engineering Physics and Computation, Technische Universität München, Boltzmannstr. 15, Garching bei München 85748, Germany.

[2] Munich Institute of Integrated Materials, Energy and Process Engineering, Technical University of Munich, Lichtenbergstr. 4a, Garching 85748, Germany.

*Corresponding author, yu.jiao@tum.de

**Abstract**
We present an efficient, fully conservative numerical scheme valid in the entire range of highly to weakly compressible flows using a single-fluid four equation approach together with multi-component thermodynamic models. The approach can easily be included into existing finite volume methods on compact stencils and enables handling of compressibility of all involved phases including surface tension, cavitation and viscous effects. The mass fraction (indicator function) is sharpened in the two-phase interface region using the algebraic interface sharpening technique Tangent of Hyperbola for INterface Capturing (THINC). The cell face reconstruction procedure for mass fractions switches between an upwind-biased and a THINC-based scheme, along with proper slope limiters and a suitable compression coefficient, respectively. For additional sub-grid turbulence modeling, a fourth order central scheme is included into the switching process, along with a modified discontinuity sensor. The proposed "All-Mach" Riemann solver consistently merges the thermodynamic relationship of the components into the reconstructed thermodynamic variables (like density, internal energy), wherefore we call them All-Mach THINC-based Thermodynamic-Dependent Update (All-Mach THINC-TDU) method. Both, liquid-gas and liquid-vapor interfaces can be sharpened. Surface tension effects are taken into account by using a Continuum Surface Force (CSF) model. In order to reduce spurious oscillations at interfaces we decouple the computation of the interface curvature from the computation of the gradient of the Heaviside function. An explicit, four stage low storage Runge-Kutta method is used for time integration.
The proposed methodology is validated against a series of reference cases, such as bubble oscillation/advection/deformation, shock-bubble interaction, a vapor bubble collapse and a multi-component shear flow. The results of a near-critical shock/droplet interaction case are superior to those obtained by WENO3 and OWENO3 schemes and support that the proposed methodology works well with various thermodynamic relations, like the Peng-Robinson equation of state. Finally, the approach is applied to simulate the three-dimensional primary break-up of a turbulent diesel jet in a nitrogen/methane mixture including surface tension effects under typical dual-fuel conditions.
The obtained results demonstrate that the methodology enables robust and accurate simulations of compressible multi-phase/multi-component flows on compact computational stencils without excessive spurious oscillations or significant numerical diffusion/dissipation.

# 1.Introduction

Simulation of flow processes in internal combustion engines (ICE) usually involves multicomponent flows with phenomena such as deformation of material interfaces, interaction with shock waves, and evaporation and condensation. Here, we focus on compressible two-phase flows with multicomponent fluids, with two-phase jet flows or droplet breakup in internal combustion engines being practical applications. Basically, there are two different numerical approaches for such flows: Methods that assume sharp interfaces and those that rely on mixture assumptions. Since the flow field and material properties are generally discontinuous across the interface between different phases or components, treating material boundaries becomes a key problem.

Sharp interfaces are generally achieved by (Interface tracking type) Front-Tracking[1,2], (Interface capturing type) Level-Set methods (LS)[3-5], Geometric interface reconstruction (GIR) based volume of fluid methods (VOF)[6-7] like Piecewise-Linear Interface Calculation (PLIC)[8] and the isoAdvector Method[9]. Front tracking attempts to combine the characteristics of both Lagrangian and Eulerian schemes to maintain a sharp boundary, where Lagrangian markers are adopted to distinguish the domains taken by different fluids. However, this is computationally expensive for complex geometries especially for strong topological deformations of the interface. A traditional Level-Set (LS) function is a signed-distance function, that represents the shortest distance to the interface, where the interface is represented as the zero contour. It is accurate in computing surface normals and curvatures with naturally smooth LS field and is suitable for determining surface tension. However, the additional steps for reinitialization and mass conservation involve additional computational cost because the LS function loses its signed distance property after the advection step and the mass of each phase is not automatically conserved. Moreover, it is expensive if the topology changes. GIR methods could also provide a clear/sharp two-phase interface, but are also computationally expensive.

Generally, diffuse interfaces are introduced by phase field models (PF)[10-13] or (interface capturing type) algebraic interface sharpening (AIS) based models, where no clear boundary between phases exists and the interface often covers more than a single cell. Phase-field methods generally adopt the convective Cahn-Hilliard or the Allen-Cahn equation to describe interfacial motions and naturally have anti-diffusive characteristics. However, these methods require considerable computational effort to resolve interfacial structures and have limited application to realistic two-phase problems in industry. Typically, the AIS model uses the Heaviside function or indicator function to distinguish different phases. The AIS method usually takes advantage of the natural conservation of mass and the effective capture of changes in interface topology, such as the breaking or reconnecting of interfaces. The various fluid components are artificially induced to mix at the unresolved interface. In this way, a thin mixing zone is created, even with non-miscible fluids. This has the great advantage that a single set of equations can be used to describe the two-phase flow properties throughout the domain, without having to explicitly track the sharp interface.

Approaches with diffuse interfaces are efficient and also suitable for three-dimensional cases with different fluid components. The corresponding diffuse interface effects are purely due to numerical processes and approximate a sharp gradient with the mesh size approaching zero. The four-equation model[14-25] usually takes the mass fraction as the Heaviside function, and the five-equation model[26] usually takes the volume fraction as the Heaviside function. It is noted that the species-mass conservation is always maintained for the five-equation model, regardless of the numerical treatment of the volume fraction of the advection equation, while the conservative form of the additional species-mass conservation equation should be carefully achieved in the fully conservative four-equation model. The mass fraction-based four-equation model can account for an additional component with an additional species-mass conservation equation, in contrast to the volume fraction-based five-equation model, which requires two equations for an additional species (advection equation for the volume fraction and equation for species mass conservation) and thus requires additional computational effort.

A well-known advantage of a mass fraction-based four-equation model (over an LS formulation) is that it can handle flows where the species front is not initially present but forms during the calculation (as in chemical reactions, cavitation, or condensation). While Gama-transport based four equation models[27](mixture of mass, momentum, energy and additional advection parameters) are ideal for stiffened gas type EOS, mass-fraction-based four equation models can be efficiently applied to real fluid models. The disadvantage of quasi-conservative models is poor conservation of species masses (the mass of each species is not accurately conserved)[28-31] and that artificial temperature spikes can occur[28].

In summary, we focus on a fully conservative four-equation model based on the mass fraction, which is easily extensible to additional species (dual-fuel conditions), suitable for real fluids, and usable for complex flows including cavitation. Since the working fluids in DFICE always contain a fraction of free gas, the proposed model is extended to include such a gas fraction in the fluid.

Artificial interface sharpening methods (AIS) strive to algebraically reconstruct the two-phase interface or modify the RHS of the governing equations to avoid excessive interfacial diffusion, at relatively low additional computational cost. Examples include the Flux-Corrected Transport (FCT) scheme[32-34], post-processing anti-diffusion methods[35,36], the Tangent of Hyperbola for INterface Capturing (THINC) method[37-39], Compressive Interface Capturing Schemes for Arbitrary Meshes (CICSAM) [40], additional (artificial) compression term approaches [41-44], bounded variation or TVD methods [45-47], ENO or WENO approaches[48,49], Multidimensional Universal Limiters with Explicit Solution (MULES)[50,51] and High Resolution Interface Capturing schemes (HRIC)[52].

Various methods have been developed to bring the numerical two-phase interface with the least dissipation into agreement with the macroscopic hypothesis of a continuous medium. For example, adjusting the limiters in the TVD-MUSCL scheme[51] or the weighting coefficients in the WENO scheme[53,54] can help reduce the numerical dissipation and avoid oscillations in the discontinuous interface region. Furthermore, the reconstruction of the parameter space near the cell surfaces($S_L$ and $S_R$) and the development of specific Riemann solvers($S^*=f(S_L,S_R)$) can reduce the numerical dissipation at interfaces.

Among all types of AIS methods, the THINC-based method is an interesting option because it generally requires less computational effort. In the THINC method presented by Xiao[37], the hyperbolic tangent function was used to evaluate the numerical flux for the advection equation of the VOF function, which aimed to calculate the moving interface algebraically (without complex geometric reconstruction). For multidimensional calculations, the numerical flux for each direction was determined by operator splitting. Then, the research focused on improving the description of the interface shape and avoiding the surface curvature. Subsequently, a Boundary Variation Diminishing (BVD) scheme[55] was proposed and the following studies[56-59] focus on selecting different reconstruction candidates to minimize jumps at cell boundaries. BVD usually reconstructs variable candidates for each cell interface and then decides on the final candidates, which results in additional computational resources and communication time for parallel computations. We focus on a robust, efficient, and compact four-cell scheme, so further details related to BVD are not part of this work.

So far, the THINC method has been developed and used for multiphase flows. However, it should be noted that THINC-based methods have been studied to a limited extent for complex compressible multicomponent applications involving shock waves, turbulence, cavitation, and moving interfaces, e.g., for high-speed atomization in a compressible multicomponent environment. Daniel[60] use THINC in MUSCL scheme for incompressible multiphase flows. To simulate a supersonic liquid jet[61], a five-equation model was combined with a volume fraction THINC when capillary forces were neglected. A five-equation model combined with a density THINC was further implemented to simulate atomization [62]. Nonomura et al. [63] developed the THINC method with two compressible fluids and sharpened the volume fraction only, while Shyue and Xiao[64] presented a single-fluid multicomponent flow model and used the reconstructed volume fraction to extrapolate other conservative parameters across the cell interface, both neglecting the surface tension.

Coupled methods such as THINC-LS[65] and THINC-LS-VOF[66] were developed to improve the interface description and numerical accuracy, but they generally require significantly more computational resources to achieve good performance. The above THINC-based method is generally used to reconstruct the volume fraction of a five-equation model. Generally such models do not directly consider a Riemann solver of the cell interface. Rather, they decouple the relationship between pressure and density as well as internal energy in the reconstruction process. In addition, a realistic-flow such as the viscous term, surface tension, gravity, real-fluid and cavitation effects are generally ignored or not fully considered.

These findings motivate the development of a numerical method that has, among others, the following properties: (1) Easy to implement; (2) Robust performance on realistic flow problems; (3) Accurate, highly efficient, and cost-effective; (4) Suitable for all Mach numbers; (5) Direct combination of turbulence models; (6) Physically consistent; (7) Flexible and scalable. The development of such an unusual method faces significant challenges, including: (1) Spurious oscillations can occur in complex two-phase simulations, especially when modeling large discontinuities; (2) The tendency to smear the interfaces is not easily reduced while maintaining high computational efficiency for three-dimensional practical problems; (3) Robustness is maintained for various types of realistic flows; (4) It is not easy to describe surface tension effects with low computational cost while reducing spurious velocities; (5) Thermodynamic consistency between variables requires consideration.

Our current work aims to solve these difficult problems, and we propose a robust FVM density-based framework with the following properties: (1) Fully conservative four-equation model with no additional advection equation for the volume fraction; (2) THINC-based reconstruction functions applied separately for the liquid-gas two-phase interface and the liquid-vapor interface; (3) Compact four-templates with low communication in parallel calculations, especially for three-dimensional flows; (4) Efficient method for describing surface tension; (5) Correction of the thermodynamic relations of a fluid for the reconstruction step (TDU): The effects of the Riemann solver are incorporated into the computational interaction process and thermodynamic consistency between variables is maintained. We propose a new four-equation (mass fraction) all-Mach THINC-TDU method using a single-fluid thermodynamic model representing a multicomponent fluid that fully accounts for viscosity and surface tension, gravity, real-fluid as well as cavitation effects. It can robustly capture two-phase interfaces (liquid-gas and liquid-vapor interfaces) while maintaining interfacial equilibrium. The discritisation scheme offers implicit subgrid-scale modelling capabilities.

## 2. Numerical model

In present work, the proposed numerical model adopts the fully conservative form of the compressible Navier-Stokes equations for two or more compressible fluids and adds additional conservation equations for the mass of (at least) one of the components. This approach is usually referred as 4 equation model [15,67], which, in our case, includes viscous effects, surface tension and gravity effects. Besides ensuring full conservation, the approach discussed in the following sections can efficiently be integrated into existing compressible finite volume methods with only minor effort due to the compact (4-cell) stencil.

## 2.1. Governing equations

The governing equations are the well known Navier-Stokes equations in conservation from for compressible fluids and fluid mixtures. q denotes mixture properties as obtained from the total mass, total momentum and total energy of the mixture and $\xi_{Gasi}$ denote the mass fraction of species "i" for a mixture of i+1 fluids. In the following, the governing equations for a compressible mixture of three fluids are presented:

$$\partial_t \boldsymbol{q} + \nabla \cdot [\boldsymbol{C}(\boldsymbol{q}) + \boldsymbol{S}(\boldsymbol{q})] = \boldsymbol{Q} \tag{2-1}$$

Where, the state vector $\mathbf{q} = [\rho, \rho u, \rho E, \rho\,\xi_{Gas1}, \rho\,\xi_{Gas2}]^T$ contain the conserved variables for density, momentum, total energy and gas species. $C_i(q)$, $S_i(q)$ and $Q$ refer to convective term, stress term and source term, respectively,

$$\boldsymbol{C}(\boldsymbol{q}) = \boldsymbol{u}\boldsymbol{q} = \boldsymbol{u}\begin{bmatrix}\rho\\\rho\boldsymbol{u}\\\rho H\\\rho\xi_{Gas1}\\\rho\xi_{Gas2}\end{bmatrix},\ \boldsymbol{S}(\boldsymbol{q}) = \begin{bmatrix}0\\p\boldsymbol{I}-\boldsymbol{\tau}\\-\boldsymbol{u}\boldsymbol{\tau}-k_c\nabla T\\0\\0\end{bmatrix}\ \text{and}\ \boldsymbol{Q} = \begin{bmatrix}0\\\rho f\\\boldsymbol{u}\rho f\\0\\0\end{bmatrix}, \tag{2-2}$$

where, I refers to the unit tensor; $\rho H = \rho E + p$ is the total enthalpy; $\tau$ is the viscous stress tensor, $\tau = \mu(\nabla u + (\nabla u)^T - 2/3(\nabla \cdot u)I)$ and $\mu$ refers to the dynamic viscosity; $k_c$ refers to the thermal conductivity; f refers to the volume force, for surface tension and gravity, $f = [f_1, f_2, f_3]^T = [\delta\sigma k n_1, \delta\sigma k n_2 + g, \delta\sigma k n_3]^T$; k is the curvature, δ refers to the Dirac function that is nonzero only on the interface, $\sigma$ denotes surface tension coefficient, n refers to the normal gradient of a indicator function; g refers to the gravity term.

## 2.2. Barotropic thermodynamics equilibrium model for coupled (one-fluid) multi-component flow

First, we extend a single-fluid model with an additional gas phase to a single-fluid multicomponent model with multiple gas components, based on prior work by Örley et al.[21] and Trummler et al.[22].
A typical three-component two-phase fluid Φ = {L, M, Gas1, Gas2} is selected, which refers to a liquid component, a liquid-vapor mixture and two gas components denoted as Gas1 and Gas2, respectively.
For the liquid component

Yu Jiao*, Steffen J. Schmidt, Nikolaus A. Adams

$$\rho_L = \rho_{sat,liq} + \frac{1}{c_L^2}(p_L - p_{sat}) \quad, p_L \geq p_{sat} \,, \tag{2-3}$$

where $\rho_{sat,liq}$ is the liquid saturation density at its saturation pressure $p_{sat}$.
For the liquid-vapor mixture

$$\rho_M = \rho_{sat,liq} + \frac{1}{c_M^2}(p_M - p_{sat}) \quad, p_M < p_{sat} \,. \tag{2-4}$$

Since $c^2 = (\partial p/\partial \rho)|_{s=const}$, the mixture speed of sound is approximated as $c_M = (p_{sat}/\rho_{sat,liq})^{1/2}$.
For the non-condensable gas phase

$$\rho_{Gasi} = \frac{p_{Gasi}}{R_{Gasi}T_{refi}} \,, \tag{2-5}$$

i.e., both gas components are treated as ideal gas, $R_{Gas1}$ and $R_{Gas2}$ refer to the specific gas constant, $T_{ref1}$ and $T_{ref2}$ refer to the corresponding temperatures.
From the volume fraction $\alpha_\Phi = V_\Phi/V$ and the mass fraction $\xi_\Phi = m_\Phi/m$ of component $\Phi$, it is obvious that $\rho_\Phi = m_\Phi/V_\Phi = \xi_\Phi m/(\alpha_\Phi V) = \xi_\Phi \rho/\alpha_\Phi$, and $\rho = m/V = \sum_\Phi \xi_\Phi \rho = \sum_\Phi \alpha_\Phi \rho_\Phi$. Naturally $\sum_\Phi \alpha_\Phi = 1$ and $\sum_\Phi \xi_\Phi = 1$, $\rho = \alpha_{L/M}\rho_{L/M} + \alpha_{Gas1}\rho_{Gas1} + \alpha_{Gas2}\rho_{Gas2}$.

$$\rho_{Gasi} = \frac{\xi_{Gasi}}{\alpha_{Gasi}}\rho = \frac{p_{Gasi}}{R_{Gasi}T_{refi}} \,. \tag{2-6}$$

Where $\alpha_{Gasi} = \xi_{Gasi}\rho R_{Gasi}T_{refi}/p_{Gasi}$. Thus $\rho = \alpha_{L/M}\rho_{L/M} + \alpha_{Gas1}\rho_{Gas1} + \alpha_{Gas2}\rho_{Gas2} = (1 - \alpha_{Gas1} - \alpha_{Gas2})\rho_{L/M} + \xi_{Gas1}\rho + \xi_{Gas2}\rho$. Finally, we obtain the coupled (one-fluid) multi-component equation of state $(1 - \sum \rho\xi_{Gasi}R_{Gasi}T_{refi}/p_{Gasi})(\rho_{sat,liq} + \frac{1}{c^2}(p_M - p_{sat})) - (1 - \sum \xi_{Gasi})\rho = 0$. It indicates $p = f(\rho, \xi_{Gasi})$ with the equilibrium assumption. If the pressure is higher than the saturation pressure of the liquid, then is no vapor. Otherwise we have

$$\alpha_V = \frac{V_{vap}}{V} = \begin{cases} (1 - \alpha_{Gas1} - \alpha_{Gas2})\frac{\rho_{sat,liq} - \rho_M}{\rho_{sat,liq} - \rho_{sat,vap}} & ,\rho < \rho_{sat,liq} \\ 0 & ,\rho \geq \rho_{sat,liq} \end{cases} \,. \tag{2-7}$$

We assume that there is no pure vapor and limit the largest volume fraction of vapor component to 99.5%. The mixtures viscosity is

$$\mu_{mix} = (1 - \alpha_{Gas1} - \alpha_{Gas2})[(1 - \alpha_v)(1 + 5/2\alpha_v)\mu_{liq} + \alpha_v\mu_{vap}] + \alpha_{Gas1}\mu_{Ga1} + \alpha_{Gas2}\mu_{Ga2} \tag{2-8}$$

The thermodynamic equilibrium model for coupled (one-fluid) multi-component flow is detailed in Appendix A.

## 2.3. Surface tension modeling

For surface tension modeling, the distribution characteristics of the indicator function or the Heaviside function (scalar in the two-phase flow) generally have a great impact on the surface tension calculation at the two-phase interface. By using various methods to sharpen the two-phase interface, the accuracy of the surface tension calculation can be improved. In addition, numerous numerical methods have been proposed to improve the balance between pressure and surface tension effects to reduce the parasitic current or velocity. For example, balance between compressive and surface tension forces can be achieved by discretizing the surface tension and compressive forces at the same location. As shown in Table2.1, we divide the surface stress modeling features into three types, according to their expressions and discrete forms. In the current work, Continuum Surface Force (CSF) model[68] is used(f), along with proper methods to obtain norm-direction gradient of scalar ($\nabla H$) and curvature($k$). Details are explained in Section3.2.

Table 2.1. Classifications for surface tension modelling methods.

| Types | Classification Criteria | Terms/ Functions | Examples | Comments |
|---|---|---|---|---|
| 1 | Heaviside function and its normal gradient | $\delta_s n = \nabla H$ | • Volume fraction function[68]<br>• Level set function[4]<br>• Dirac delta of ghost fluid method (GFM)[69,70] | Obtain the unit normal across the interface |
| 2 | Methods for curvature calculation | $k$ | • Smoothed volume fraction method[68]<br>• Height function method[71-73]<br>• Smooth the original curvature through kernel[74]or weight coefficient method[75] | Obtain accurate curvature |
| 3 | Methods for Surface Tension Force discretization (explicit curvature maybe not required) | f | • Continuum surface force (CSF)[68], in some cases the curvature and norm-direction gradient of scalar are decoupled (like GFM method or level set method)<br>• Continuum surface stress(CSS)[76,77]<br>• Balanced Continuum surface force (bCSF) method or ghost fluid models[78]<br>• Improved Riemann solver[79,80]<br>• Sharp Surface Force (SSF) model[81,82]<br>• Independent Source item effects methods[83] | Surface tension acts on pressure field but not the velocity field, thus the parasitic currents can be reduced |

## 2.4. Turbulence modeling

For turbulence modeling, a high order improvement of THINC-TDU is proposed. To model the effects of sub-grid turbulence in the current four equation model, an implicit large eddy simulation (iLES) method is used. The iLES approach for compact stencils proposed by Egerer et al.[23] was based on the Adaptive Local Deconvolution Method (ALDM) by Adams et al.[84] and Hickel et al.[85,86] to deal with resolved sub-grid turbulence. The truncation error of the discretization scheme is learned from the data to serve as a sub-grid scale (SGS) model for turbulence. Hickel et al.[86,87] also developed a compressible version of ALDM, with shock capturing abilities while smooth pressure waves and turbulence are propagated without excessive numerical dissipation. More details related to ALDM are given in the work of Hickel[85,88,89] and Egerer et al.[23]. The ALDM method has been used by Örley[21,90], Trummler[22,91] in the two phase flow, which proves that it is a proper method in solving the fully compressible two phase turbulent flow. Based on previous work and in order to include the turbulence modeling, a high order improvement of THINC-TDU is proposed, in which the ALDM reconstruction is implemented to the momentum equation.

Here we briefly summary the iLES method for a scalar nonlinear transport equation,

$$\frac{\partial u}{\partial t} + \frac{\partial}{\partial x} f(u) = 0 \tag{2-9}$$

And a linear low-pass filter operation could be represented as[86]

$$\bar{u}(x) = \int_{-\infty}^{+\infty} G(x - x') u(x') dx' = G * u \tag{2-10}$$

By projecting the filtered continuous function onto the numerical grid $x_N = \{x_i\}$, the LES discretization of the transport equation is obtained as

$$\frac{\partial \bar{u}_N}{\partial t} + G * \partial_x f_N(u_N) = \varepsilon_{sgs} \tag{2-11}$$

And the sub-grid scale error/residual arises from the nonlinearity of f(u) and is obtained as $\varepsilon_{sgs} = G * \partial_x f_N(u_N) - G * \partial_x f_N(u)$, in which the inverse-filter operation $u_N = G^{-1} * \bar{u}_N$.

A finite-volume discretization could be presented as

$$\frac{\partial \bar{u}_N}{\partial t} + G * \partial_x \widetilde{f_N}(\widetilde{u_N}) = 0 \tag{2-12}$$

where $\bar{u}_N$ presents the approximately deconvolved parameter, and the top-hat filter kernel G is

$$G(x - x_i) = \begin{cases} 1/\Delta x, & |x - x_i| \leq \Delta x/2, \\ 0, & \text{otherwise.} \end{cases} \tag{2-13}$$

A modified differential equation (MDE) can be formulated as [84]

$$\bar{u}(x) = \int_{-\infty}^{+\infty} G(x - x') u(x') dx' = G * u \tag{2-14}$$

$$\frac{\partial \bar{u}_N}{\partial t} + G * \partial_x f_N(u_N) = \varepsilon_N \tag{2-15}$$

Where $\varepsilon_N = G * \partial_x f_N(u_N) - G * \partial_x \widetilde{f_N}(\widetilde{u_N})$ presents the truncation error owning to the spatial discrete format.

When the truncation error reproduces the physical properties of the exact sub-grid scale error/residual, we call the numerical discretization physically consistent. In this case, the numerical truncation error functions as sub-grid scale model and we call the numerical scheme with $\varepsilon_N$ an iLES model. In other words, if $\varepsilon_N$ approximates $\varepsilon_{sgs}$ in some sense for finite $\Delta x$ we obtain an implicit SGS model contained within the discretization.

# 3. Numerical methods

## 3.1. Overall description of All-Mach THINC-TDU algorithm

Based on the finite volume method CATUM[21-23,90-91], we propose the following numerical scheme where surface tension effects, a modified discontinuity sensor and real fluid effects are newly considered. In addition, numerical dissipation reduction and thermodynamic consistency at interfaces are achieved.

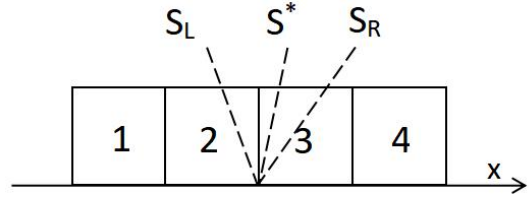

Fig.3.1 Schematic diagram of four-cell stencil

As shown in Fig.3.1, the following algorithm is based on a compact four-cell stencil, but it can be extended to a higher order with a larger stencil. Here, we give this easy-to-implement four-cell stencil algorithm to solve the proposed model including viscous effects, surface tension effects, and gravity effects. The thermodynamic model used in the following part is related to the barotropic thermodynamic equilibrium model for coupled (single-fluid) multicomponent flows proposed in 2.2. It can be extended to consider internal energy and real fluid effects, see Appendix B.

The procedures of the All-Mach THINC-TDU method, within one Runge-Kutta time integration sub-step are presented in **Algorithm 1**. An explicit, four stage low storage Runge-Kutta method is used for time integration. Appendix D describes details of sub-grid turbulence model, as an extension.

**Algorithm 1:** All-Mach THINC-TDU algorithm including liquid-vapor and liquid-gas interface sharpening

| | |
|---|---|
| **For** | each stage k(1~4) of Runge-Kutta scheme **do** |
| | // Application of (All-Mach) Riemann solver; |
| | // Reconstruction of cell interface variables with sharpened liquid-gas interface; |
| | // Sharpening liquid-vapor interface; |
| | // Discretization of surface tension; |
| | // (Optional)High order improvement with turbulence modeling(Appendix D); |
| | // Calculation of convective fluxes, viscous fluxes, sources terms(gravity); |
| | // Update conservative variables $q^k \rightarrow q^{k+1}$; |
| | // Update primitive variables $\Phi^k \rightarrow \Phi^{k+1}$ and thermodynamics parameters according to EOS; |
| | // Update boundary conditions. |
| **End** | |

In the following, we present the "all-Mach" Riemann solver and introduce the reconstruction procedure for sharpening the liquid-gas interface. We also specifically propose procedures for sharpening the liquid-vapor interface, and describing surface tension discretization as well as high order improvement with turbulence modeling, finally give the flowchart of the newly developed algorithm.

## 3.2. Description of Riemann solver

To overcome the low Mach number problem, we use the "all-Mach" interface pressure reconstruction (asymptotically consistent pressure flow definition) and the modified numerical flux of Schmidt et al. [92].

$$p^* = \frac{p_L + p_R}{2} \tag{3-1}$$

The following Riemann solver is applied for the interface velocity $S^* = u^*$, which determines the upstream direction.

$$u^* = \frac{\rho_L c_L u_L + \rho_R c_R u_R + p_L - p_R}{\rho_L c_L + \rho_R c_R} \tag{3-2}$$

## 3.3. Reconstruction of interface variables

At distinct interfaces between a liquid and a gaseous component or mixture we use THINC to reconstruct the mass fraction for the interface free-gas component(s) along with TDU to consistently reconstruct related thermodynamic quantities. Regions that do not contain (sharp) interfaces are treated as single fluid and MUSCL type reconstructions along with proper limiters are used.

Let $\xi_{Gas}$ denote a scalar quantity, such as the mass fraction of a gas component. Let $\xi_{Gas}^1$ through $\xi_{Gas}^4$ be the cell averaged scalars in cells 1 to 4 as sketched in figure 3.1.

• **Discontinuous interface between cells 2 and 3**

In order to compute the flux across the cell interface between cells 2 and 3 we reconstruct left and right side values of the scalar $\xi_{Gas}$ as

$$\xi_{Gas}^L = min(\xi_{Gas}^1, \xi_{Gas}^3) + \frac{max(\xi_{Gas}^1, \xi_{Gas}^3) - min(\xi_{Gas}^1, \xi_{Gas}^3)}{2}\left(1 + \theta \frac{tanh(\beta) + E}{1 + E \cdot tanh(\beta)}\right), \quad (3\text{-}3)$$

where, $\theta = \begin{cases} 1 & if\ \xi_{Gas}^3 \geq \xi_{Gas}^1 \\ -1 & otherwise \end{cases}$, $E = \frac{exp(\theta\beta(2F-1))/cosh(\beta) - 1}{tanh(\beta)}$, $F = \frac{\xi_{Gas}^2 - min(\xi_{Gas}^1, \xi_{Gas}^3) + \gamma}{max(\xi_{Gas}^1, \xi_{Gas}^3) + \gamma}$;

$$\xi_{Gas}^R = min(\xi_{Gas}^2, \xi_{Gas}^4) + \frac{max(\xi_{Gas}^2, \xi_{Gas}^4) - min(\xi_{Gas}^2, \xi_{Gas}^4)}{2}(1 + \theta E), \quad (3\text{-}4)$$

where, $\theta = \begin{cases} 1 & if\ \xi_{Gas}^4 \geq \xi_{Gas}^2 \\ -1 & otherwise \end{cases}$, $E = \frac{exp(\theta\beta(2F-1))/cosh(\beta) - 1}{tanh(\beta)}$, $F = \frac{\xi_{Gas}^3 - min(\xi_{Gas}^2, \xi_{Gas}^4) + \gamma}{max(\xi_{Gas}^2, \xi_{Gas}^4) + \gamma}$).

$\gamma$ is a small number to avoid division by zero and the parameter $\beta = 1.6$ was found to give suitable results. The literature[37,55-60,65,66,93-98] discusses values for β in the range between 1.6 and 3.

• **Regions without discontinuous interfaces**

In smooth regions, an All-Mach MUSCL type reconstruction with proper limiters is used. Free-gas mass fractions are described here to illustrate the process. They are discretized by a second order upwind biased reconstruction with proper slope limiter.

In order to compute the flux across the cell interface between cells 2 and 3 we reconstruct left and right hand values of the scalar $\xi_{Gas}$ as $\xi_{Gas}^L = \xi_{Gas}^2 + 1/2 f(r_{i+1/2}^-)(\xi_{Gas}^2 - \xi_{Gas}^1)]$ and $\xi_{Gas}^R = \xi_{Gas}^3 - 1/2 f(r_{i+1/2}^+)(\xi_{Gas}^4 - \xi_{Gas}^3)]$. $f(r)$ represents the slope limiter function that the ratio of upwind to central differences could be adjusted according to various limiters, for example,

·Minmod limiter: $f(r) = max(0, min(1, 2r), min(r, 2))$.
·Koren limiter 3rd-order accurate for smooth data: When r>0, $f(r) = min(2.0, 2.0 * r, (1.0 + 2.0 * r)/3.0)$; otherwise $f(r) = 0$.

The slope function is decided according to the upwind direction. If upwind direction is in the positive direction, $r_{i+1/2}^- = (\xi_{Gas}^2 - \xi_{Gas}^1)/(\xi_{Gas}^3 - \xi_{Gas}^2)$, otherwise, $r_{i+1/2}^+ = (\xi_{Gas}^4 - \xi_{Gas}^3)/(\xi_{Gas}^3 - \xi_{Gas}^2)$.

Together with following TDU idea, the All-Mach THINC-TDU method is obtained.

• **TDU idea**

Other interfacial parameters are updated based on the interfacial pressure and interfacial gas mass fraction according to the thermodynamic equilibrium function, so that the numerical discrete format of the indicator matches the discrete density format and thermodynamic consistency is maintained

$$\rho^* = f(p^*, \xi_{Gas1}^*, \xi_{Gas2}^*) \quad (3\text{-}5)$$

Another option is to reconstruct the density and then update the mass fraction according to function $\xi_{Gas1}^* \& \xi_{Gas2}^* = f(p^*, \rho^*)$. Then we obtain the volume fraction according to thermodynamic relations

$$\alpha_{Gas1}^* \& \alpha_{Gas2}^* = f(\rho^*, p^*, \xi_{Gas1}^*, \xi_{Gas2}^*) \quad (3\text{-}6)$$

## 3.4. Sharpening the liquid-vapor interface

Moreover, in cases of cavitation/bubble collapse with liquid and liquid-vapor components, it is difficult to sharpen the liquid-vapor two-phase interface directly because the volume fraction of vapor is determined according to the barotropic relations in Section 2.2, while in cases with liquid and non-condensable gas components, the gas-liquid two-phase interface is sharpened according to the above steps. Here, the procedure for sharpening the liquid-vapor two-phase interface is further elaborated.

**Remark 1.** For regions where liquid-vapor two phase interface are detected by a sensor, or a discontinuous liquid-vapor interface region meets the requirements $\varepsilon < \alpha < 1 - \varepsilon$ and $(\overline{\alpha_i} - \overline{\alpha_{i-1}})(\overline{\alpha_{i+1}} - \overline{\alpha_i}) > 0$, where ε is a small positive parameter comparing with $\overline{\alpha_i}$, the density is constructed by the THINC-based idea,

$$\rho^L = min(\rho^1, \rho^3) + \frac{max(\rho^1,\rho^3)-min(\rho^1,\rho^3)}{2}(1+\theta\frac{tanh(\beta)+E}{1+E\cdot tanh(\beta)}) \qquad , \qquad (3\text{-}7)$$

herein, $\theta = \begin{cases} 1 & if\ \rho^3 \geq \rho^1 \\ -1 & otherwise \end{cases}$, $E = \frac{exp(\theta\beta(2F-1))/cosh(\beta)-1}{tanh(\beta)})$, $F = \frac{\rho^2-min(\rho^1,\rho^3)+\gamma}{max(\rho^1,\rho^3)+\gamma})$,

$$\rho^R = min(\rho^2, \rho^4) + \frac{max(\rho^2,\rho^4)-min(\rho^2,\rho^4)}{2}(1+\theta E) \qquad , \qquad (3\text{-}8)$$

where, $\theta = \begin{cases} 1 & if\ \rho^4 \geq \rho^2 \\ -1 & otherwise \end{cases}$, $E = \frac{exp(\theta\beta(2F-1))/cosh(\beta)-1}{tanh(\beta)})$, $F = \frac{\rho^3-min(\rho^2,\rho^4)+\gamma}{max(\rho^2,\rho^4)+\gamma})$. $\gamma$ is a small number to avoid division by zero and the parameter $\beta = 1.6$ was found to give suitable results.

Subsequently, according to the thermodynamic relations (2-4) and (2-7), the vapor volume fraction is obtained. Other parameters like the speed of sound need also to be updated accordingly. Modified sensors can be used to detect liquid-vapor two phase regions, which are introduced in Appendix D.

**Remark 2.** For complex cases with liquid, liquid vapor, non-condensable gas, firstly, THINC is applied for mass fraction of non-condensable gas of liquid-gas two phase interface $\xi^*_{Gas1}, \xi^*_{Gas2}$. Secondly, TDU is applied to obtain density of liquid-gas two phase interface, i.e. $\rho^* = f(p^*, \xi^*_{Gas1}, \xi^*_{Gas2})$; then THINC is applied for the density of liquid-vapor two phase interface $\rho^*$ (**Remark 1**). Finally, TDU is applied to obtain vapor of liquid-liquid vapor two phase interface $\alpha^*_v = f(p^*, \rho^*, \xi^*_{Gas1}, \xi^*_{Gas2})$ as well as other parameters like speed of sound.

## 3.5. Discretization of surface tension

The calculation of the normal gradient and curvature depends on which surface tension model is used. Here we decouple the calculation of the gradient of the Heaviside approximation from the curvature calculation. We use a Gaussian convolution kernel (filter) to smooth this indicator function, and then use the smoothed parameters for the curvature calculation(along with appropriate discrete method with weight factors). This simple method attempts to strike a balance between robustness and time overhead in all-Mach THINC-based interface compression.

$$\text{2D Gaussian convolution kernel (filter)}\ \frac{1}{16}\begin{bmatrix} 1 & 2 & 1 \\ 2 & 4 & 2 \\ 1 & 2 & 1 \end{bmatrix} \qquad (3\text{-}9)$$

As shown in Fig.3.2, this 2D kernel is adapted for three dimensional cases, where the normalization coefficient is 1/64,

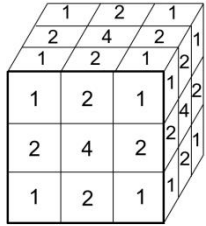

Figure.3.2. 3D Gaussian convolution kernel blurring filters (with the normalization coefficient 1/64).

After the smoothing step, the curvature is obtained from

$$k = \nabla\cdot(\frac{\nabla\tilde{\alpha}}{|\nabla\tilde{\alpha}|}) \qquad (3\text{-}10)$$

For the calculation of the curvature, the discrete shape of the gradient part can follow the Gaussian divergence theorem or the following method can be used.

As shown in Fig. 3.3, the curvature is calculated through adjacent 9 cells in 2D (27 cells are used for the 3D case). The gradient in each cell interface is first calculated repeatedly in x and y directions. Then the gradient of the four vertices [i+1/2,j+1/2], [i+1/2,j-1/2], [i-1/2,j+1/2], [i-1/2,j-1/2] is calculated. The last step consists in finding the gradient of the middle grid [i, j] by averaging the gradient of the smoothed volume fraction in the four vertices [i+1/2,j+1/2], [i+1/2,j-1/2], [i-1/2,j+1/2], [i-1/2,j-1/2]. The weights of the smoothed volume fraction of the neighboring cells are implicitly considered and act like the effects of a Gaussian convolution kernel (filter).

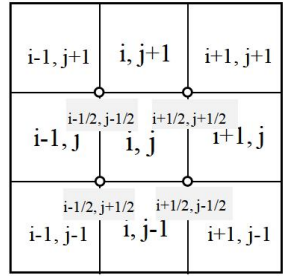

Figure 3.3. The diagram to calculate curvature

As explained in Section 2.3, there are some other options/methods to describe surface tension effects in a surface tension dominant flow, and most of them could give even better results when combined with the adaptive mesh refinement (AMR) method.

For example, the height function (HF) method with $3\times3\times3$, $3\times3\times5$, or $3\times3\times7$ cells for the 3D case and $3\times3$, $3\times5$, or $3\times7$ cells for the 2D case; can provide better accuracy but is more time-consuming. With a large number of templates requires special treatment (template reduction), which reduces its robustness. Here we give only expressions that can be directly combined with the current scheme: $k = -\nabla\cdot\mathbf{n} = -H''(x_0)/[1+H'(x_0)^2]^{3/2}$, where, $\mathbf{n} = -\frac{1}{[1+H'(x_0)^2]^{1/2}}\begin{pmatrix}H'(x_0)\\1\end{pmatrix}$, $H_i = \sum_{j-3}^{j+3}\alpha_{i,j}\cdot\Delta y_{i,j}$, $H'_{(x_0)} = (H_{i+1} - H_{i-1})/2\Delta x$, $H''_{(x_0)} = (H_{i+1} + H_{i-1} - 2H_i)/(\Delta x)^2$. $\Delta x$ and $\Delta y$ are the mesh size in x and y direction respectively.

An associated Riemann solver with the model for flows with interfaces including capillary effects was proposed[79], by replacing the pressure jump condition across contact discontinuities, $[P] = 0$, by $[P] = -\sigma k[\alpha]$, $\Delta P^* = -\sigma k\Delta\alpha$, $s^* = [p_R - p_L + \rho_L u_L(s_L - u_L) - \rho_R u_R(s_R - u_R) - \sigma k(\alpha_R - \alpha_L)]/[\rho_L(s_L - u_L) + \rho_R(s_R - u_R)]$. Therefore, the original Riemann solver (such as HLLC) could be improved to balance the surface force well and reduce the parasitic currents. Another possibility would be the CSS model, which uses the surface tension at the cell surface and can be cast in conservative form, but also tends to generate spurious currents. The performance of the current surface tension modeling will be discussed in details.

## 3.6. Flowchart of All-Mach THINC-TDU algorithm

Fig.3.4 shows the flowchart of the newly developed algorithm and briefly summarizes the content of the whole work. We implement a fully conservative form of the four-equation model for the compressible single-fluid multicomponent flow, which includes the thermodynamic relation between single-fluid and multicomponent flow. The THINC-based reconstruction method is combined with the correction of the thermodynamic relations in the reconstruction step with the All-Mach Riemann solver(TDU). Mass fraction and density as well as the process of internal energy reconstruction are associated with the same thermodynamic relations (EOS). After applying THINC for the mass fraction reconstruction step for liquid-gas cases, the All-Mach-Riemann solver is used in combination with the EOS to update the mass fraction primitive parameters (such as density, internal energy). In this way, these parameters (density and internal energy) are automatically sharpened, maintaining thermodynamic consistency and reducing numerical error by using the same numerical scheme. This All-Mach THINC thermodynamic-dependent update (All-Mach THINC-TDU) method preserves physical compatibility and is conservative. It sharpens the two-phase interface (including the liquid-vapor region and the liquid-gas region). An advection equation for the volume fraction is not required since it can be obtained from thermodynamic relations.

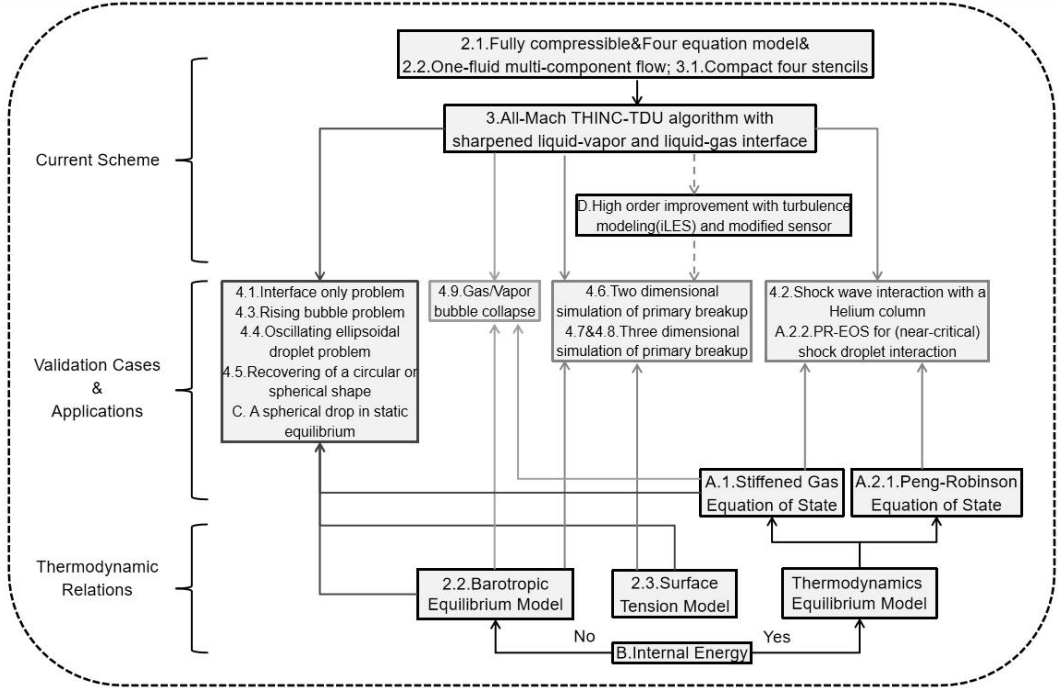

Figure.3.4. Flow chart of current numerical scheme.

# 4.Validations and results

## 4.1. Interface only problem

The "interface only problem" is used to verify the applicability of the method for simulating two-phase flows without spurious pressure/velocity oscillations or strong interface smearing. In the following, the one-dimensional advection of a two component fluid, representative for convection-dominant two-phase flows, is investigated. Initial and boundary conditions are shown in Fig.4.1. A computational domain of length 0.2m is discretized with 100 cells in x-direction. Here, we use the thermodynamic closure relations presented in equations (A.1.1)~(A.1.5) with parameters including initial pressure and velocity listed in Table 4.1.

Table 4.1. Thermodynamics parameters of current case.

| Component | $\gamma$ | R (kJ/(kg·K)) | $C_v$ (kJ/(kg·K)) | $P_\infty$ (Pa) | T (K) | $\rho$ (kJ/m$^3$) | $P_0$ (Pa) | U (m/s) |
|---|---|---|---|---|---|---|---|---|
| 1 | 5 | 7500.75 | 1875.18 | 0 | 300 | 4.444 | $10^7$ | 5 |
| 2 | 1.6 | 283.33 | 472.22 | 0 | 300 | 117.647 | $10^7$ | 5 |

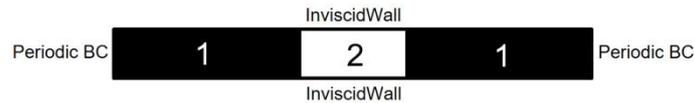

Figure 4.1 Interface only problem: the 1D advection of a square column

For the All-Mach MUSCL-TVD-TDU scheme, limiters are used to control the reconstruction of the primitive variables at cell interfaces. As shown in Table 4.2, we compare various limiters and discuss their suitability in the following section. Koren's limiter is used for velocity and various limiters are applied to reconstruct the mass fraction. Density and pressure at cell interfaces are obtained as described in Section 3.2~3.4.

Yu Jiao*, Steffen J. Schmidt, Nikolaus A. Adams

Table 4.2 Limiters for scalar transportation.

| Name | Expressions |
|---|---|
| | $r_i = \dfrac{\overline{\Phi_i} - \overline{\Phi_{i-1}}}{\overline{\Phi_{i+1}} - \overline{\Phi_i}}$ |
| VanAlbada's limiter[101] | $f(r) = (r + r*r)/(1 + r*r)$ |
| VanLeer's limiter[102] | $f(r) = (r + abs(r))/(1.0 + abs(r))$ |
| MinMod limiter[99] | $f(r) = \max(0, \min(1, r))$. |
| Chatkravathy limiter[103] | $f(r) = \max(0.0, \min(1.0, 4r))$ |
| Monotonized Central (MC) limiter[104] | $f(r) = \max(0.0, \min(2, 2r, 0.5(1 + r)))$ |
| Koren's limiter[100] | $f(r) = \max(0, \min(2.0, 2.0*r, (1.0 + 2.0*r)/3.0)$ |

Fig.4.2 compares the volume fractions after four flow through times (at t=0.16s) as obtained by the investigated limiters and by the proposed sharpening methodology "THINC-TDU". Obviously, the phase interface shows significant smearing for most of the limiters, although Korens' limiter gives superior results. However, the proposed methodology outperforms all others noticeably as shown in Fig.4.2.

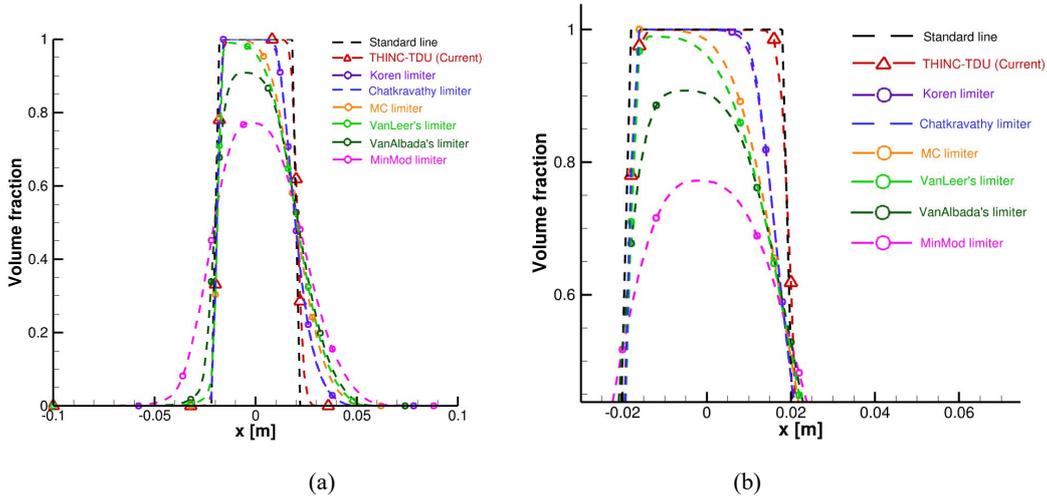

(a)      (b)

Figure 4.2. Effects of different flux limiters on the 1D interface advection in the convective dominant two phase flow (a) volume fraction distribution using All-Mach MUSCL-TVD-TDU method with different limiters or All-Mach THINC-TDU method (b) zoomed view of (a), (c)volume fraction distribution after 4 times periodic flow, at T=0.16s, by MinMod liomiter.

As shown in Fig.4.3(a), using the All-Mach THINC-TDU method, the distribution of the phase interface remains sharp over time. Fig.4.3(b)~Fig.4.3(c) show that pressure as well as velocity maintain in their initial condition even after four periodic flow through times, which proves that the method can keep interface sharp and avoid pressure or velocity oscillations.

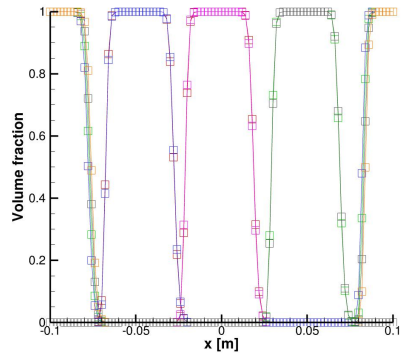

(a) Two phase interface evolution T=0.1s, 0.2s, 0.3s, 0.4s, 0.6s, 1.0s, 1.3s, 1.4s, 1.5s, 1.6s.

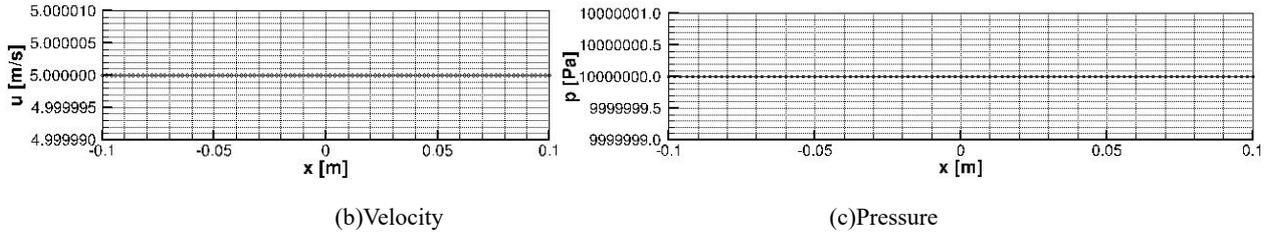

(b) Velocity  (c) Pressure

Figure 4.3 Velocity and pressure and volume fraction distribution after 4 times periodic flow, at T=0.16s, by All-Mach THINC-TDU.

## 4.2. Shock wave interaction with a Helium cylinder

In this part we assess the ability of the proposed methodology to capture the shock-wave bubble interaction processes on example of a the well-known test-case specified in[105]. The computational domain and the initial conditions are shown in Figure 4.4 and summarized below. A uniform mesh with $14400 \times 1600$ cells in stream-wise and normal direction (0.055625mm resolution) are adopted. The upper and lower boundary of the computational domain are inviscid solid walls while on the left and right boundary zero gradients for the flow variables[105,106] are prescribed. Table 4.4 shows the fluid properties and their thermodynamic modeling. The thermodynamic closure relations used in this case are given in equations (A.1.1)~(A.1.5) of Appendix A.

During the evolution process, the two phase interface is kept sharp by the proposed methodology.

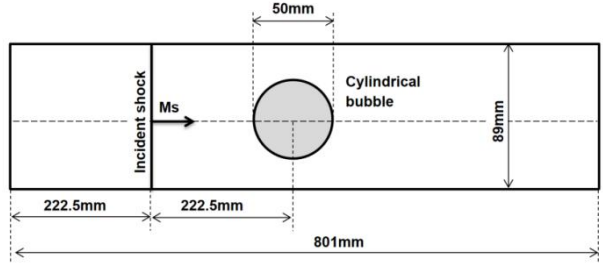

Figure 4.4. The schematic of shock-bubble interaction computational domain (sketch map).

The specific conditions are presented in Table4.3, which correspond to the non-dimensionalized initial conditions in[106]

Table 4.3. Initial conditions for shock bubble interaction case.

| Stage | $p$ [Pa] | $u$ [m/s] | $v$ [m/s] | $\rho$ [kg/m$^3$] | Non-dimension ($p,u,v,\rho$) |
|---|---|---|---|---|---|
| pre-shocked air | 101325 | 0 | 0 | 1.225 | (1,0,0,1) |
| post-shocked air | 159059 | 113.5 | 0 | 1.686 | (1.5698,-0.394,0,1.3764) |
| helium | 101325 | 0 | 0 | 0.169 | (1,0,0,0.138) |

Table 4.4. Gas properties adopted in the shock-bubble interaction simulations[105].

| Gas Component | $\gamma$ | R [kJ/(kg·K)] | Cv [kJ/(kg·K)] | Ms | Mesh number |
|---|---|---|---|---|---|
| air | 1.400 | 0.287 | 0.720 | 1.22 | 14400×1600 (0.055625mm) |
| He | 1.670 | 2.080 | 3.110 | | |

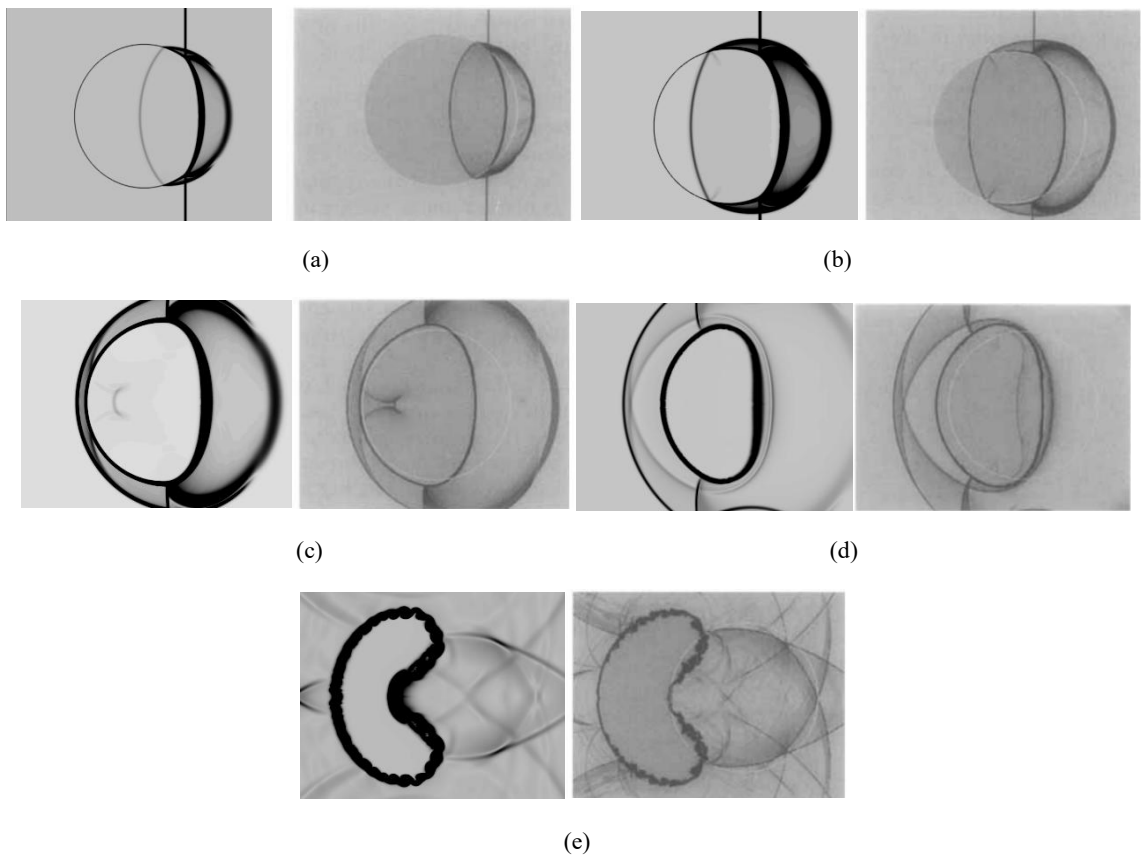

Figure 4.5. Numerical schlieren images for the evolution of shocked air-helium interaction. (a)32μs, (b)52μs, (c)72μs, (d)102μs, (e)245μs; the first column presents current numerical results, while others show results from reference.

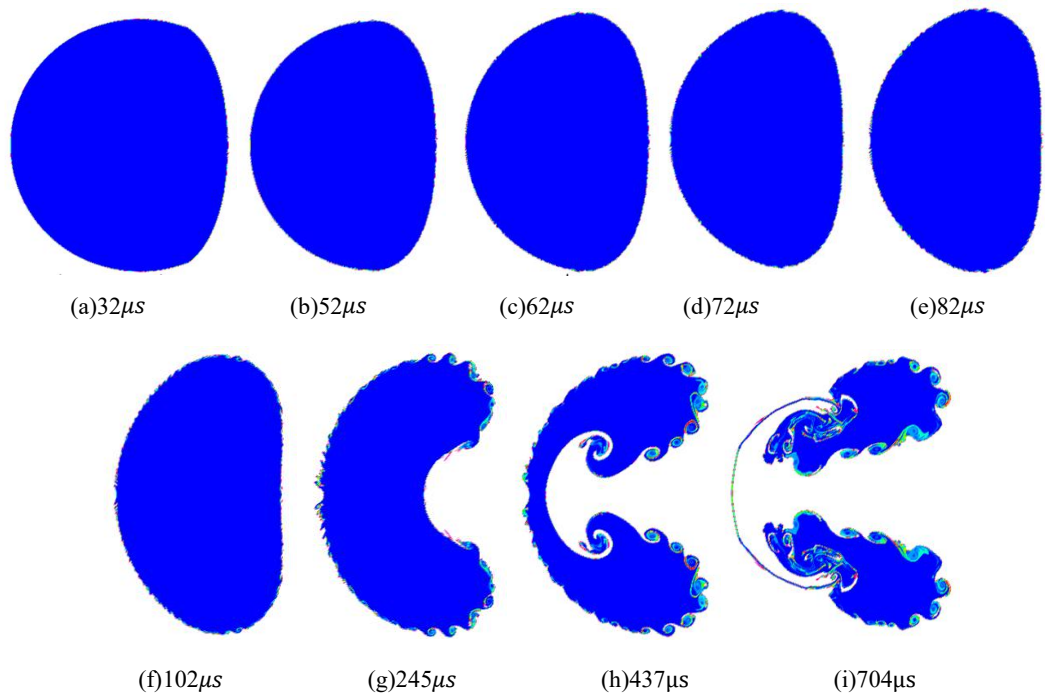

Figure 4.6. Evolution of volume fraction from the interaction of shocked air-helium interaction. (a)32μs, (b)52μs, (c)62μs, (d)72μs, (e)82μs, (f)102μs, (g)245μs, (h)437μs, (i)704μs.

As shown in Fig.4.5, we can see that the results of the helium-air test-case are in good agreement with the results from Quirk and Karni [105]. Highly detailed structures are obtained and shown in Fig.4.7. The methodology clearly resolves the physical phenomena while ensuring robust performance.

For the idealized Schlieren images, the following form to set pseudo-schlieren values is adopted,

$$\emptyset = exp\left(-C\frac{|\nabla\rho|+A}{B+A}\right) \tag{4-1}$$

Here, $\emptyset$ refers to the pseudo-schlieren value, $|\nabla\rho| = [(\partial\rho/\partial x)^2 + (\partial\rho/\partial y)^2]^{1/2}$, and the three values A,B,C can be adjusted according to display effects. Generally, A=0 and $B = |\nabla\rho|_{max}$ and it decays to $\emptyset = exp(-C|\nabla\rho|/|\nabla\rho|_{max})$. The displayed gray scales are adjusted according to the method recommended in the reference[105].

The evolution time steps selected are most close to the reference time, so small deviation are acceptable since results of the reference time could not be perfectly found.

As shown in Fig.4.7, the evolution history of three characteristic two phase interface points are also in very good agreement with Quirk and Karni[105].

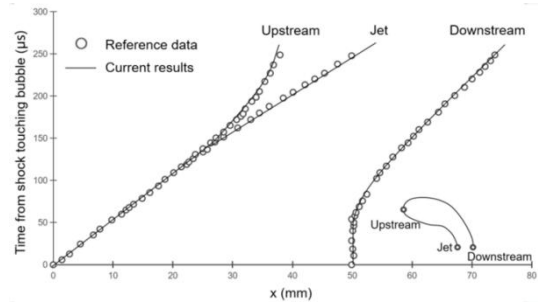

Figure 4.7 Space-time diagram for three characteristic two phase interface points in the shock-bubble problem, comparing with results from Quirk and Karni.

## 4.3. Rising bubble

In order to validate the proposed methodology with respect to effects caused by surface tension, viscous forces as well as gravity, we simulate the classical rising bubble case[107] with a mesh resolution of 100×200 cells for Case1 and 160×320 cells for Case2 in x- and y-direction. The computational domain is shown in Fig.4.8. As illustrated in Table 4.4, two different cases are investigated. The thermodynamic closure relations presented in equations (2-3)~(2-8) are used.

Table 4.4. parameters for two rising bubble cases.

| | $\rho_1$ (kg/m³) | $\rho_2$ (kg/m³) | $\mu_1$ (Pa·s) | $\mu_2$ (Pa·s) | g(m/s²) | σ(N/m) |
|---|---|---|---|---|---|---|
| Case1 | 1000 | 100 | 10 | 1.0 | 0.98 | 24.5 |
| Case2 | 1000 | 1.0 | 10 | 0.1 | 0.98 | 1.96 |

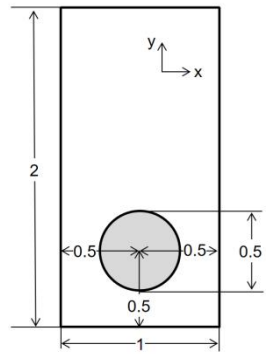

Figure 4.8. Computation domain of rising bubble test case.

In order to demonstrate the significant improvement of the proposed methodology compared to previous work we first show results obtained with the All-Mach MUSCL-TVD-TDU in Fig.4.9(a) and Fig. 4.10(a). As one can see, the two phase interface heavily smears out, although positions and shapes show reasonable agreement with the references given in Fig.4.9(c) and Fig.4.10(c)[108,109], respectively. In contrast, the proposed All-Mach THINC-TDU method produces excellent results as shown in Fig.4.9(b) and Fig. 4.10(b) [108,109]. Thus, the applicability of our numerical scheme for the cases including low Mach numbers, surface tension, viscous effects, large density rations as well as gravity is demonstrated.

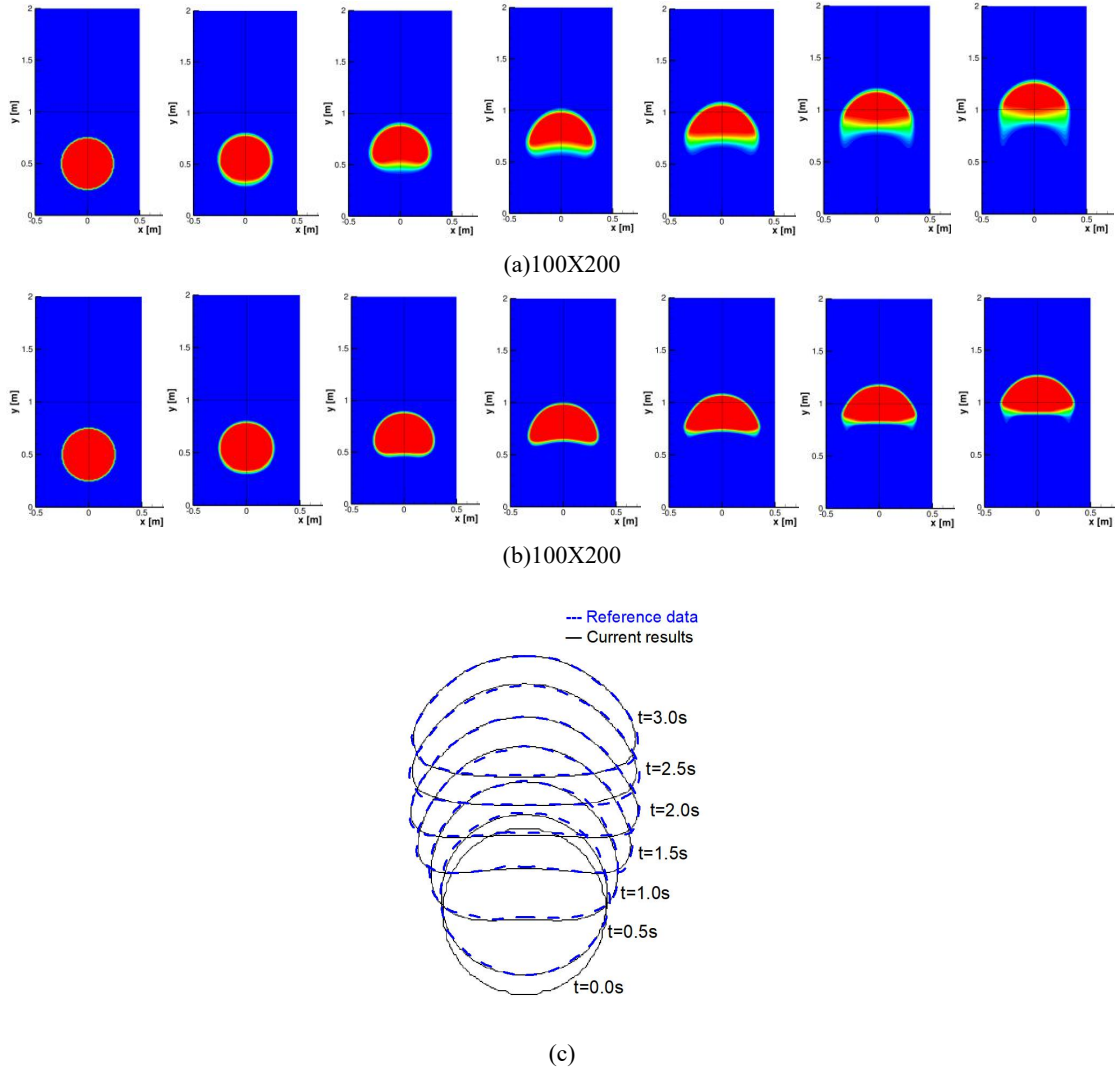

Figure 4.9. Rising bubble evolution of case1 at t=0s, 0.5s, 1s, 1.5s, 2s, 2.5s, 3s (from left to right) (a)Results of All-Mach MUSCL-TVD-TDU method, (b)Results of All-Mach THINC-TDU, (c)Comparison of results between reference[109]and All-Mach THINC-TDU.

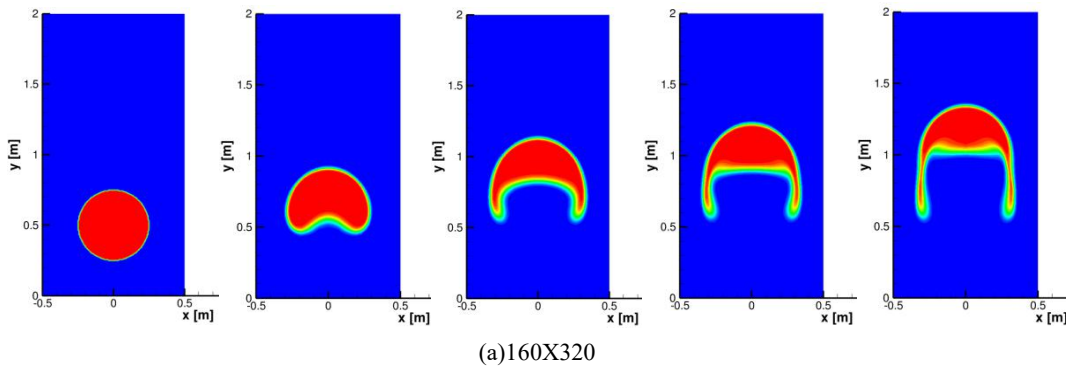

(a)160X320

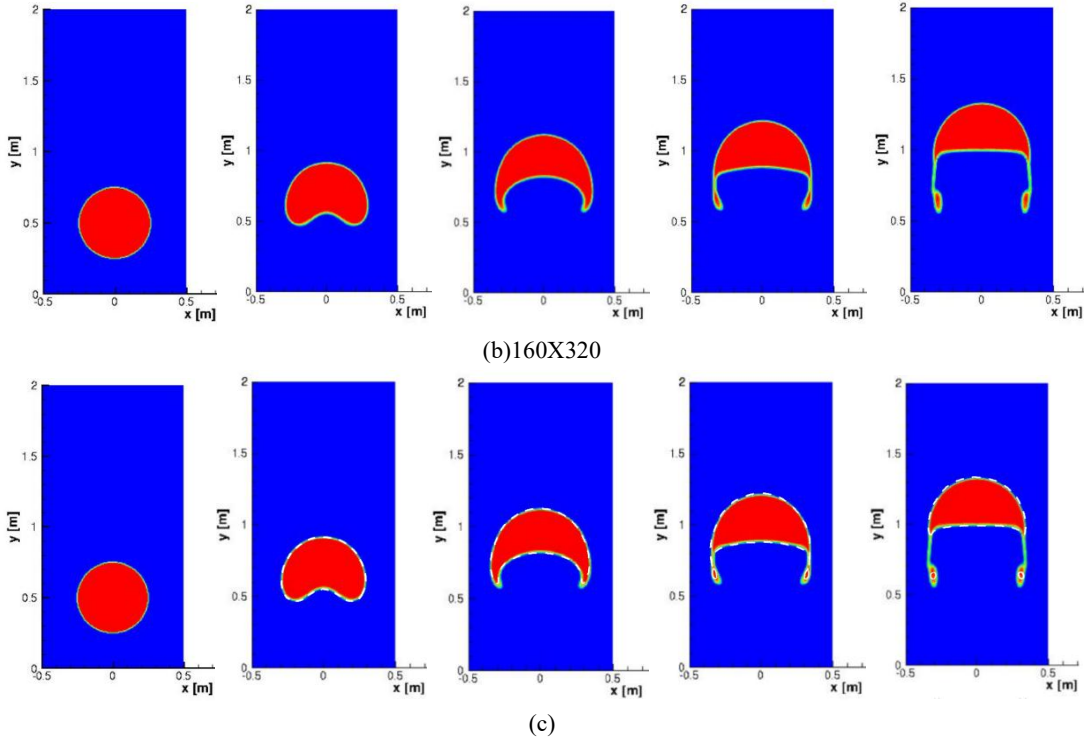

(b)160X320

(c)

Figure 4.10. Rising bubble evolution of case2, at t=0s, 1s, 2s, 2.4s, 3s (from left to right),(a)All-Mach MUSCL-TVD-TDU method, (b)All-Mach THINC-TDU results, (c)Comparison of results between Reference marked with dotted line[108] and All-Mach THINC-TDU.

## 4.4. Oscillating ellipsoidal drop

An oscillating ellipsoidal drop is used to verify the ability of the method to predict the dynamics of surface tension. We use the thermodynamic closure relations presented in equations (A.1.1)~(A.1.5) with parameters shown in Table 4.5,

Table 4.5. Thermodynamics parameters of current case.

| Component | γ | R (kJ/(kg·K)) | $C_v$ (kJ/(kg·K)) | $P_\infty$ (Pa) | T (K) | ρ (kJ/m$^3$) | $P_0$ (Pa) | $U_0$ (m/s) |
|---|---|---|---|---|---|---|---|---|
| 1 | 5 | 7500.75 | 1875.18 | 0 | 300 | 4.444 | $10^7$ | (0,0) |
| 2 | 1.6 | 283.33 | 472.22 | 0 | 300 | 117.647 | $10^7$ | (0,0) |

The quadratic computational domain is $(1.58\times10^{-3})\times(1.58\times10^{-3})$ m² which is discretized with 158×158 square cells. Initial pressure is $P_0 = 10^7 Pa$. Non-reflective boundary conditions are used. The original bubble shape is

$$\frac{(10000x-7.9)^2}{4^2} + \frac{(10000y-7.9)^2}{3^2} = 1 \quad (4-2)$$

We compare our results with an analytic expression for the oscillation period of the liquid droplet given by Fyfe et al.[110,111]:

$$T = 2\pi\sqrt{\frac{(\rho_1+\rho_2)(ab)^{3/2}}{48\sigma}} \quad (4-3)$$

In our case, the densities are 117.647 kg/m³ and 4.444kg/m³, respectively. The initial ellipsoid has short axis a=6×10$^{-4}$ m and major axis b=8×10$^{-4}$ m, surface tension coefficient is σ = 80N/m, thus a periodic time is T ≈ $2 \times 10^{-5}$s is obtained[110-114].

In Fig4.12, the oscillating bubble shape evolution are shown. The bubble shape changes into a circular shape at times T2=5×10$^{-6}$s and T4=1.5×10$^{-5}$s, while at T5=2×10$^{-5}$s the ellipsoidal shape is recovered. Figure 4.13 it shows the temporal evolution of the semi-length of the minor axis, from which an oscillation period of T=2×10$^{-5}$s is found in accordance to the analytic expression.

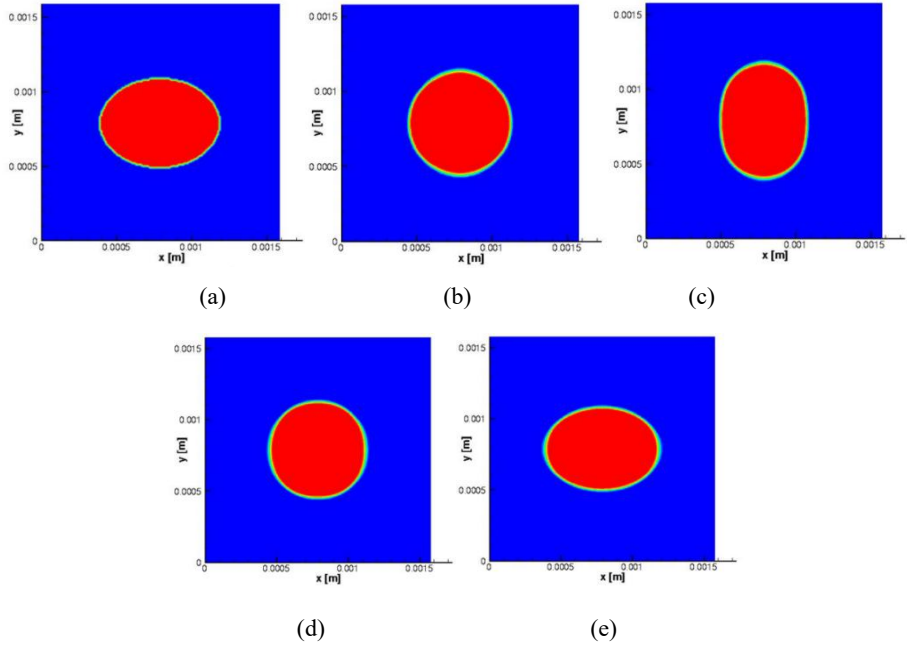

Figure 4.12 Oscillating bubble shape at (a)T1=0s, (b)T2=5×10⁻⁶s, (c)T3=10⁻⁵s, (d)T4=1.5×10⁻⁵s (e)T5=2×10⁻⁵s.

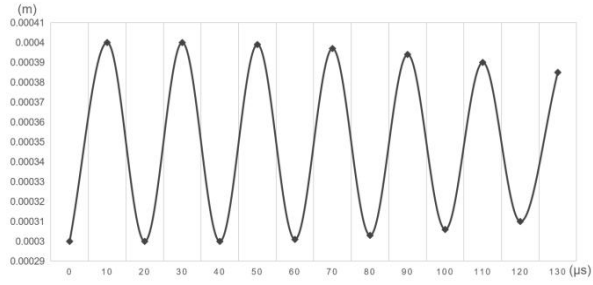

Figure 4.13. The time dependent semi-length of minor axis.

After an initial transient we compute averaged pressure fields inside and outside of the bubble and obtain a value of 241017Pa. Since the equivalent radius of the ellipsoidal bubble is R=0.346 mm and the surface tension coefficient is 80N/m,

$$[P] = \frac{\sigma}{R} \tag{4-4}$$

the theoretical value of the pressure jump for a spherical bubble is about 231214Pa, corresponding to a relative pressure error of about 4.24%, which is acceptable for the given mesh resolution.

## 4.5. Recovery of circular or spherical shape

In this subsection we investigate a two-dimensional and a three-dimensional transition from a quadratic or cubic "bubble" towards its circular or spherical shape. The two-dimensional case is identical to the one presented in the reference[115], where the computational domain is 0.75m×0.75m with mesh resolution of 150×150 cells in x-direction and y-direction, respectively. A square bubble is centred in the computational domain with an initial side length L=0.2m. The surface tension coefficient is 800N/m and the density ratio between the liquid bubble and the gaseous ambient is 1000, with $\rho_{gas} = 1\text{kg/m}^3$ and $\rho_{liquid} = 1000\text{kg/m}^3$. Non-reflective boundary conditions are used. Viscous effects as well as gravity are neglected. For this test-case, all substances are modeled by barotropic thermodynamic relations according to equations (2-3) till (2-8), where,

$\rho_{sat, liq} = 1000\text{kg/m}^3, p_{sat} = 2340, T = 336.9\text{K}, c = 1500\text{m/s}, p_0 = 1\text{bar}, R_{Gas} = 296.8, k_{Gas} = 1.4$

Our results presented in Fig.4.15(a) are in very good agreement with the reference shown in Fig.4.15(b)

[115]. Note that, contrary to the reference, we show the volume fraction while "magnified schlieren images of the mixture density" are shown in the reference but a description of the numerical method to generate schlieren images is missing.

We extend this two-dimensional recovery case to three dimensions. As shown in Fig.4.15(c)~(f), a spherical bubble evolves from an initially cubic "bubble" under the effects of surface tension, which proves the ability of current methodology to predict three-dimensional flow physics including surface tension effects.

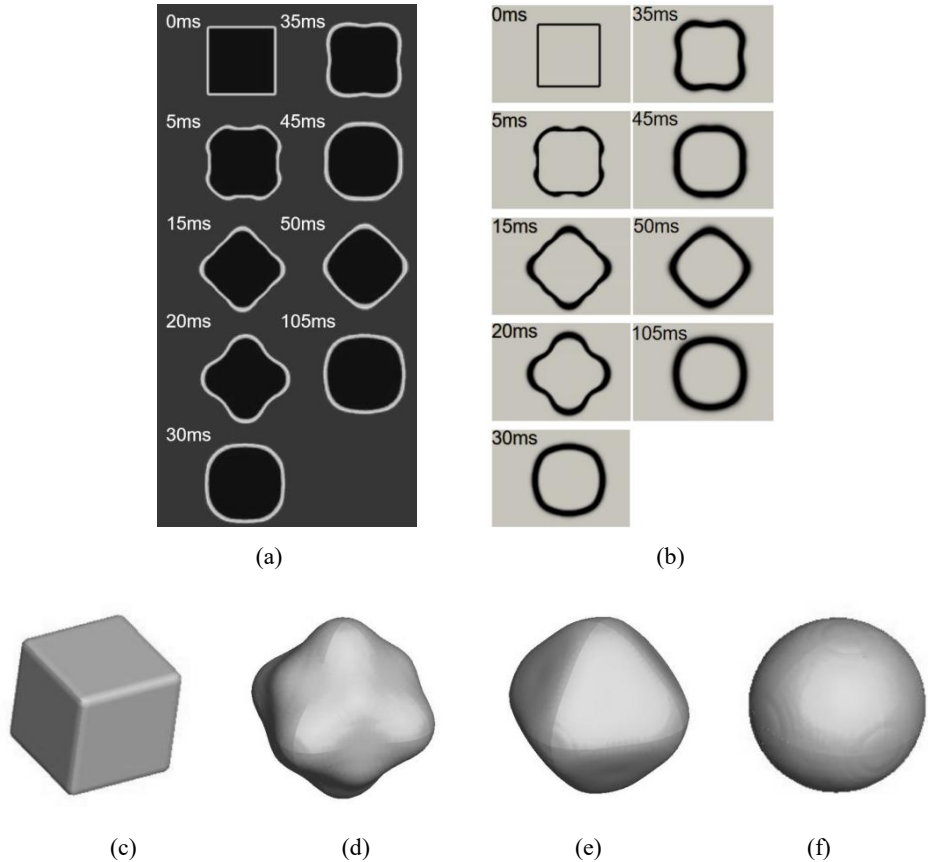

(a)  (b)

(c)  (d)  (e)  (f)

Figure 4.15.Bubble shape evolution using volume fraction contour (a) current two dimensional recovery bubble case with uniform mesh resolution 0.005m and surface tension coefficient is 800N/m, (b) Reference of two dimensional recovery bubble case with uniform mesh resolution 0.005m and surface tension coefficient is 800N/m[115],(c)~(f) evolution of three dimension recovery bubble case.

## 4.6. Two dimensional simulation of primary breakup

We apply the proposed algorithm to one of our target applications, which cover dual fuel internal combustion engine (DFICE) flow physics. In order to demonstrate the benefits of the interface sharpening approach, we simulate a two-dimensional planar shear layer under typical DFICE conditions. Figure 4.16 shows the numerical domain and the boundary conditions. The characteristic length is D=89.4 μm and the grid resolution is 0.75 μm. Identical to the upcoming chapter 4.7, the fluid properties correspond to the "SprayA-210675" test-case [116], where liquid n-Dodecane and a gas mixture of 20% Methane and 80% Nitrogen enter the domain from left, separated by a viscous wall. Symmetry boundary conditions at the top and at the bottom surfaces are prescribed. The initial chamber and ambient pressure is 60MPa. The inlet velocity of the liquid is 500 m/s and the velocity of the gas mixture is 450 m/s. For this test-case, all substances are modeled by barotropic thermodynamic relations according to equations (2-3) till (2-8).

In Fig.4.17 we compare the predicted evolution of the two-phase interface and its sharpness using our standard model (a and c, left) and the recently developed sharpening approach (b and d, right). The later one leads to a significant improvement in the prediction quality, preventing the interface from getting smeared and allowing for higher details, such as the liquid tip. Since the numerical complexity of both approaches compares well, the improved interface quality either allows for a reduction in mesh resolution or in a gain in quality.

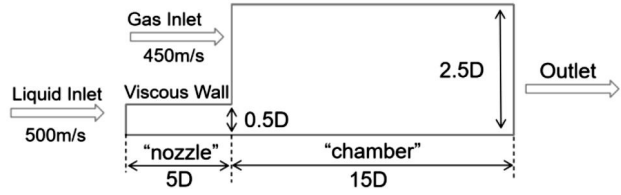

Figure 4.16. Sketch Map of "SprayA- 210675 model" Benchmark Case.

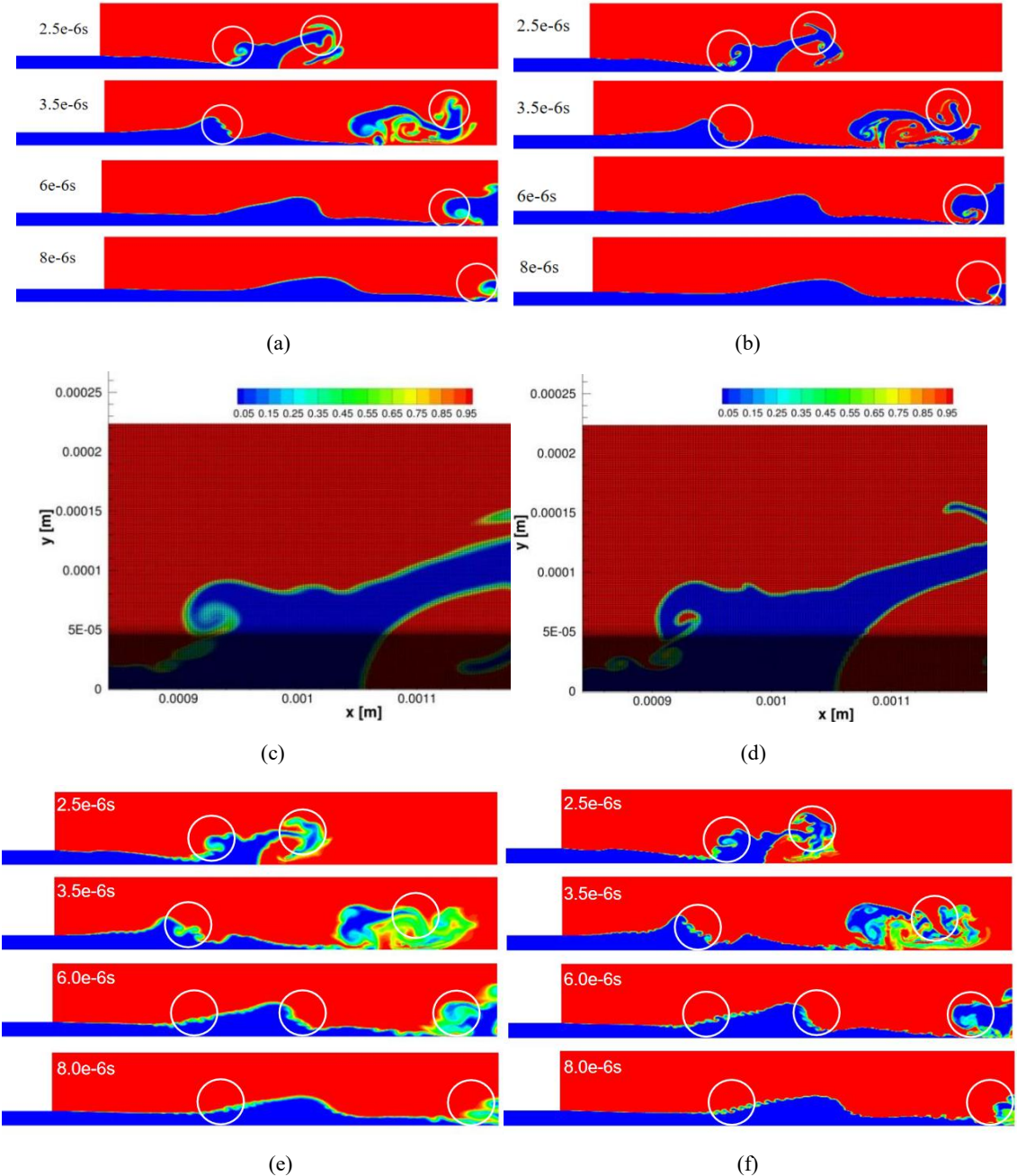

Figure 4.17. Two phase interface evolution of 2D "SprayA-210675 model" Benchmark Case (60Mpa, ΔV=50m/s), All-Mach MUSCL-TVD-TDU: (a) and (c), All-Mach THINC-TDU: (b) and(d), Original iLES (e), and high order THINC-TDU-iLES :(f)

Moreover, according to Section 2.4 and Appendix D, we compare the results of our original iLES scheme and the newly developed THINC-TDU high-order iLES algorithm. It is evident that the improved iLES scheme yields sharper two-phase interfaces. Moreover, the viscous sublattice effects (structures) in Fig.4.17(e)(f) obtained with the iLES scheme are more obvious than those in Fig.4.17(a)(b) obtained with the scheme without SGS model (Section 4.6). In Sections 4.7 and 4.8, we will show three-dimensional cases simulated with the THINC-TDU and THINC-TDU-iLES high-order algorithms.

## 4.7. Three dimensional simulation of primary breakup in dual-fuel conditions by THINC-TDU scheme

This test-case demonstrates the ability and robustness of the two-phase interface treatment methodology to be applied in engineering applications such as a high speed liquid jet discharging into a dual-fuel ambient. The test-case is commonly referred as "SprayA-210675"[116]. A block-structured o-grid with a total number of 55 million cells at a resolution of minimum 0.04μm at a time-step of 0.1 nanoseconds is used. As shown in Fig.4.18, the computational domain is 10D×10D×20D (D=89.4 μm ) in x/y/z-direction, respectively. The initial chamber and ambient pressure is 60MPa and the discharge velocity of the jet is 500 m/s. The liquid jet consists of n-Dodecane and the gas mixtures includes 20% Methane and 80% Nitrogen. For this test-case, all substances are modeled by barotropic thermodynamic relations according to equations (2-3) through (2-8). Due to the high inertia of the jet and as our focus is on the robustness of the methodology, we neglect surface tension and gravity.

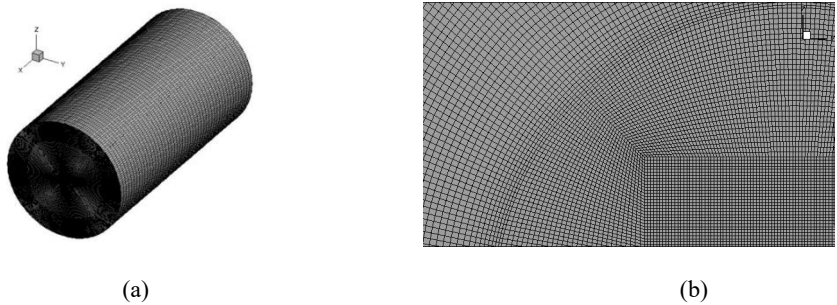

(a)  (b)

Figure 4.18: Computational domain for "Spray A-210675 model" (a) and detail of the o-grid (b).

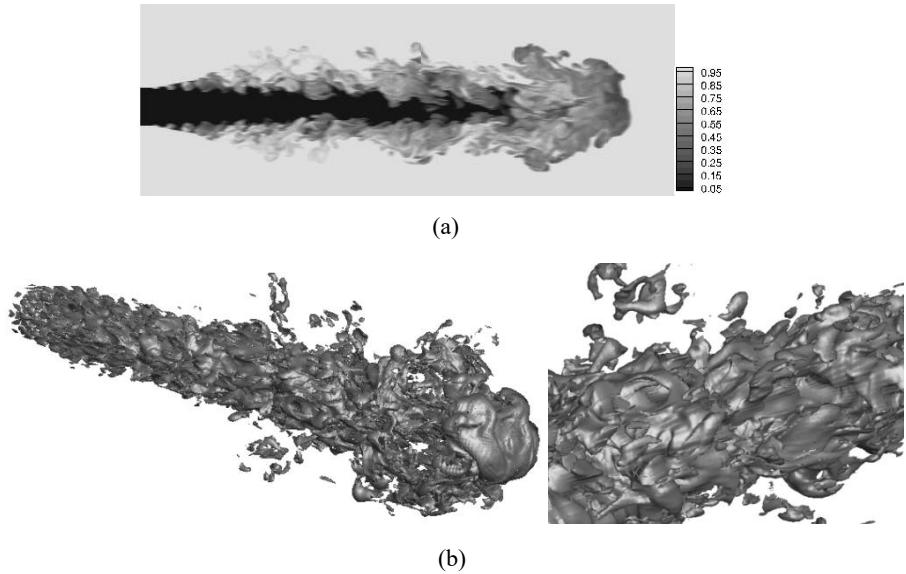

(a)

(b)

Figure 4.19. Volume fraction of Mixture gas in stream-wise middle plane with interface sharpening at T= 4.79026×10$^{-6}$s, (a) contour of volume fraction (b) ISO-surface of 50%

In Fig. 4.19, the mushroom head shape, bridges, lobes, droplets, and other typical structures can be seen in this high-resolution simulation, showing the robust performance of our method. This three-dimensional simulation represents an under-resolved DNS of high-speed primary breakup, with inherent numerical dissipation keeping the computation stable. In this case, the compression method is applied only to the region of the two-phase interface, so some nearly pure fluid regions still exhibit a tendency to smear, but it is minor.

Yu Jiao*, Steffen J. Schmidt, Nikolaus A. Adams

## 4.8. Three dimensional simulation of turbulent jet flow in dual-fuel conditions

In this section we extend the two dimensional case of Fig.4.16 to a three dimensional counterpart shown in Fig.4.21 by applying periodic boundary conditions in span-wise direction. The smallest mesh size near the wall is $0.1\,\mu m$, which meets the requirement $y^+ < 1$ based on liquid diesel properties and the relative velocity between liquid and gas. We compare the three dimensional results from the original iLES and the current high order THINC-TDU-iLES algorithm(Appendix D) along with surface tension. The surface tension coefficient of diesel adopted is 0.028N/m. The initial pressure and the ambient pressure is 6 MPa, lower than case of Section 4.7, while other parameters are identical. Actually, the effects of surface tension can be ignored in this convectional-dominant flow with large Weber number (in the region near the nozzle exit). However, we includes viscous effects as well as surface tension in order to show the performance and robustness of the current scheme.

The development of the shear layer is shown in Fig.4.22(a). It is obvious in Fig.4.22 that the improved iLES method provides sharper interfaces than the original one and the primary breakup is well captured.

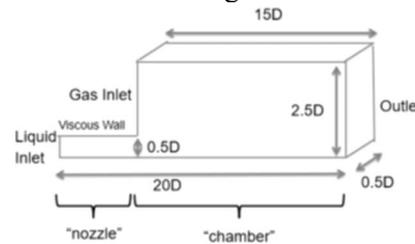

Figure.4.21. Sketch Map of "SprayA- 210675 model" three dimensional Benchmark Case.

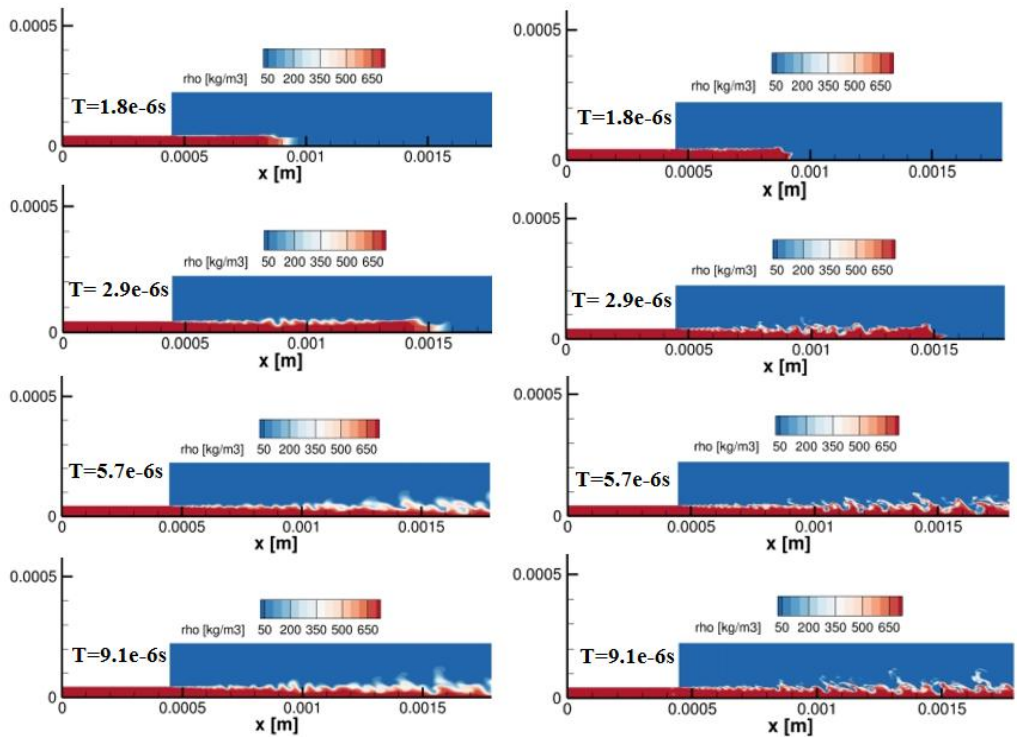

(a)Original iLES: left column, and high order THINC-TDU-iLES algorithm: Right column T=1.8×10⁻⁶s, 2.9×10⁻⁶s, 5.7×10⁻⁶s, 9.1×10⁻⁶s.

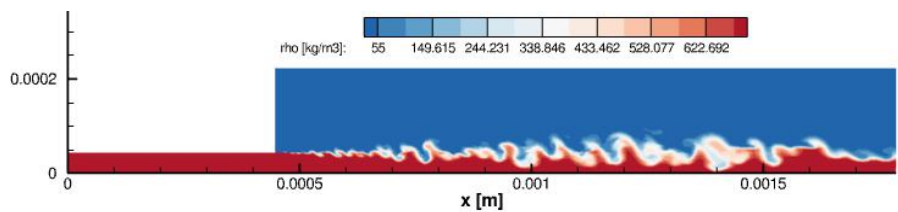

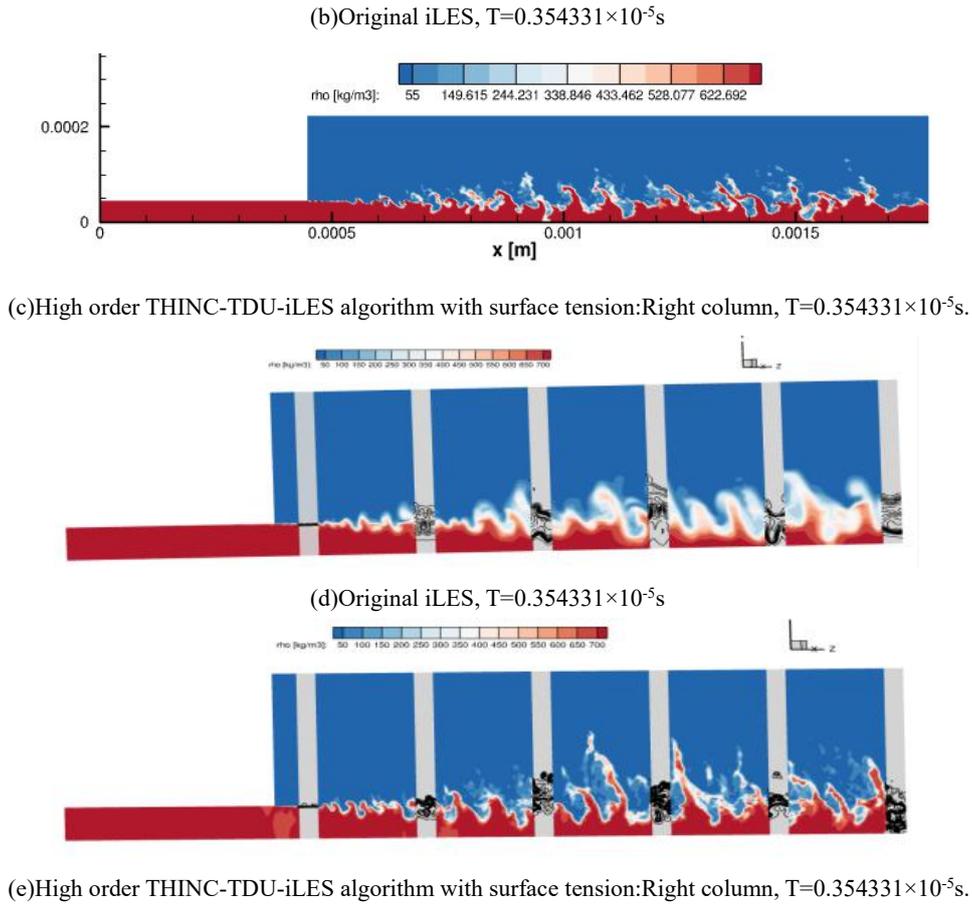

(b)Original iLES, T=0.354331×10⁻⁵s

(c)High order THINC-TDU-iLES algorithm with surface tension:Right column, T=0.354331×10⁻⁵s.

(d)Original iLES, T=0.354331×10⁻⁵s

(e)High order THINC-TDU-iLES algorithm with surface tension:Right column, T=0.354331×10⁻⁵s.

Figure.4.22. 2D density contour (a), and 3D density contour (b)~(e).

## 4.9. Gas/Vapor bubble collapse with interface sharpening

In the following, vapor and gas bubble collapses are simulated using the method of interface sharpening and TDU. In this simulation, we use the thermodynamic relations from section 2.2 and Appendix A. All configuration settings such as initial pressure distribution and initial bubble position are similar to[117]. As shown in Fig.4.23, the distance between the gas or vapor bubble and the wall is H=440μm and the radius of the bubble is R=400 μm. As shown in Fig 4.24, the meshes near the bubble are refined and the mesh size is 4μm. Around the far-field region, the meshes are enlarged. For the gas bubble we use the initial gas pressure of 3000 Pa and for the vapor bubble we assume a vapor pressure of 1342 Pa. From Fig.4.24, Fig.4.25, and Fig.4.26, it can be seen that the present scheme is suitable for bubble collapses with a sharp two-phase interface between liquid and gas or (condensable) vapor, which allows for more complicated applications such as turbulent cavitating liquid jets into a gas environment.

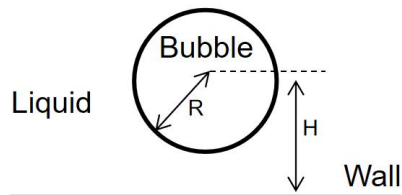

Figure.4.23. Sketch Map of bubble collapse.

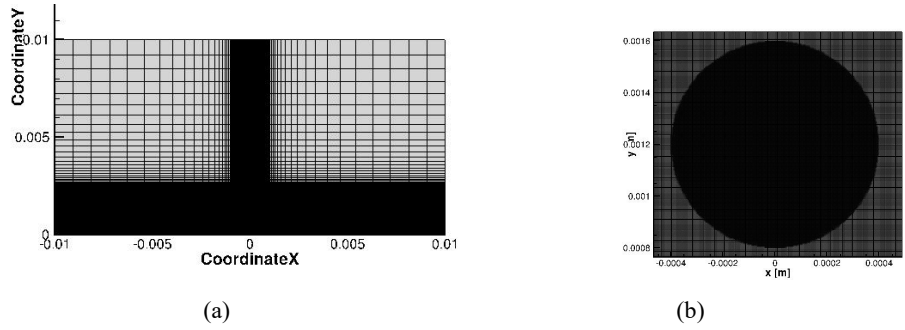

(a)                      (b)

Figure.4.24. Sketch Map of bubble collapse mesh (a), enlarged mesh around bubble (b) with smallest mesh resolution 4μm.

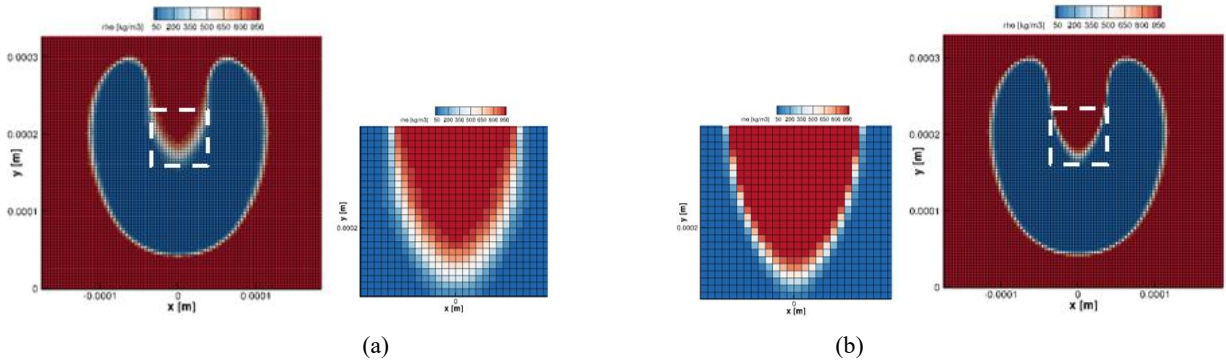

(a)                      (b)

Figure.4.25. Gas Bubble collapse with smallest mesh resolution 4μm: density contour of All-Mach MUSCL-TVD-TDU method:(a); density contour of All-Mach THINC-TDU:(b).

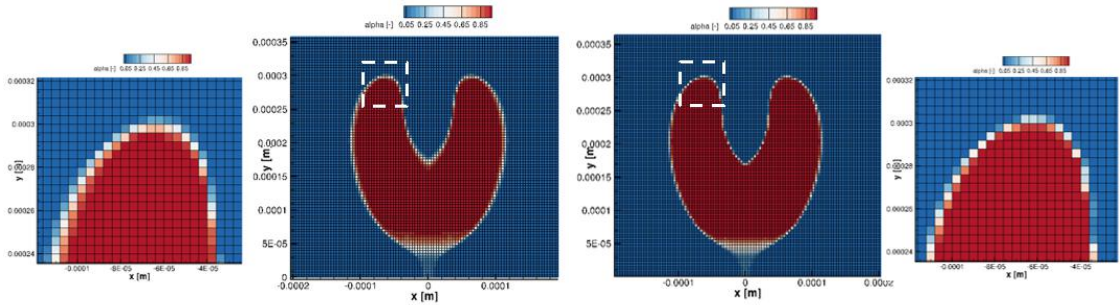

(a)Vapor contour of All-Mach MUSCL-TVD-TDU method:Left, vapor contour of All-Mach THINC-TDU:Right.

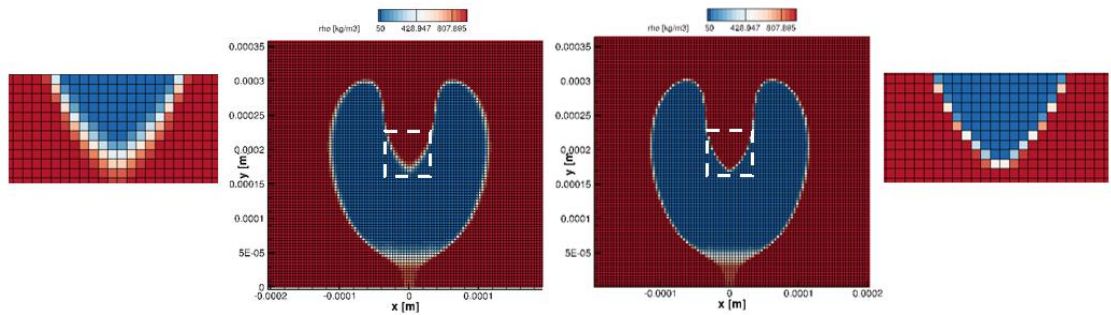

(b)Density contour of All-Mach MUSCL-TVD-TDU method:Left, Density contour of All-Mach THINC-TDU:Right.

Figure.4.26. Vapor Bubble collapse with smallest mesh resolution 4μm.

## 5.Conclusions

In this work we propose a robust four equation model using one-fluid multi-component thermodynamics relations as well as an All-Mach number consistent THINC-TDU method, which prevents two-phase interfaces from smearing. Surface tension effects, viscous effects, gravity effects, as well as shock-wave phenomena have been assessed and the results are in very good agreement with well-known reference results. Our simulation of a liquid jet in a dual-fuel environment demonstrates the suitability of the methodology to complex real-world engineering applications. An efficient implementation of the methodology into existing MUSCLE or WENO-type compressible finite volume methods on block-structured meshes is presented. A significant improvement of predicted details in compressible two-phase flows is reached while the additional computational costs are negligible. This is achieved by combining an algebraic sharpening method with a thermodynamically consistent correction procedure in the flux computation without the need of complex geometric reconstruction approaches. The approach has been implemented into our in-house code CATUM validated against a series of references and has extended to be suitable for sub-grid turbulence modelling. A shock-droplet test-case in near critical conditions with a real fluid model show that current results are superior to WENO3-JS and OWENO3 schemes (if using the same mesh resolution). In addition to gas bubble collapse, vapor bubble collapse was also performed to prove that the two-phase interface between liquid and vapor can also be sharpened using the current scheme. This demonstrates the ability of the current scheme to handle cavitation-related cases containing both condensable vapor and non-condensable gas, such as atomization using cavitation nozzles. The simulation of a three-dimensional turbulent jet flow with surface tension and viscous effects also demonstrates the high performance of the current scheme.

# Appendix A. Thermodynamics equilibrium model for coupled (one-fluid) multi-component flow

### A.1. Stiffened Gas Equation of State

In this part we take internal energy into account. The variables of the mixtures in the one-fluid multi-component model are shown in Table.A.1. The detailed prove of thermodynamic equilibrium closure could be observed in references[124-126]. It is noted that Noble-Abel Stiffened-Gas equation of state degenerates to the Stiffened-Gas equation of state by setting the covolume of the fluid b=0.

Table.A.1 variables of the mixtures in the one-fluid multi-component model

| Variables | Descriptions |
|---|---|
| Volume fractions | $\alpha_{Gas1} + \alpha_{Gas2} + \alpha_L = 1$ |
| Mass fractions | $\xi_{Gas1} + \xi_{Gas2} + \xi_L = 1$, $\xi_{Gas1} = \alpha_{Gas1}\rho_{Gas1}/\rho$, $\xi_{Gas2} = \alpha_{Gas2}\rho_{Gas2}/\rho$, $\xi_L = \alpha_L\rho_L/\rho$ |
| Pressure | $P = (\alpha_{Gas2} + \alpha_{Gas2} + \alpha_L)P = (\xi_{Gas1} + \xi_{Gas2} + \xi_L)P$ |
| Density | $\rho = \alpha_{Gas1}\rho_{Gas1} + \alpha_{Gas2}\rho_{Gas2} + \alpha_L\rho_L = (\xi_{Gas1} + \xi_{Gas2} + \xi_L)\rho$ |
| Specific total energy | $E = e + |u^2|/2$ |
| Internal energy | $\rho e = \alpha_{Gas1}\rho_{Gas1}e_{Gas1} + \alpha_{Gas2}\rho_{Gas2}e_{Gas2} + \alpha_L\rho_L e_L = \rho\xi_{Gas1}e_{Gas1} + \rho\xi_{Gas2}e_{Gas2} + \rho\xi_L e_L$ |
|  | $e = \xi_{Gas1}e_{Gas1} + \xi_{Gas2}e_{Gas2} + \xi_L e_L$ |
| Specific total enthalpy | $H = E + P/\rho = e + |u^2|/2 + P/\rho$ |
| Specific enthalpy | $\rho h = \alpha_{Gas1}\rho_{Gas1}h_{Gas1} + \alpha_{Gas2}\rho_{Gas2}h_{Gas2} + \alpha_L\rho_L h_L = \xi_{Gas1}\rho h_{Gas1} + \xi_{Gas2}\rho h_{Gas2} + \xi_L\rho h_L$ |
| Mixed viscosity | $\mu_{mix} = (1 - \alpha_{Gas1} - \alpha_{Gas2})[(1 - \alpha_v)(1 + 5/2\alpha_v)\mu_{liq} + \alpha_v\mu_{vap}] + \alpha_{Gas1}\mu_{Ga1} + \alpha_{Gas2}\mu_{Ga2}$ |

Here we use "stiffened Gas" EOS for liquid and two gas components,

$$P = (\gamma - 1)\rho(e - q) - \gamma P_\infty \qquad (A.1.1)$$

where, $\gamma$ is the heat capacity ratio $C_P/C_V$. In this way, $\rho_L = (P_L + \gamma_L P_{\infty,L})/((\gamma_L - 1)(e_L - q_L))$. Gas fluid could decay to ideal gas with $P_{\infty,Gas}=0$. Generally, for ideal gas q=0, $P_{Gas} = (\gamma_{Gas} - 1)\rho_{Gas}e_{Gas}$ and $\rho_{Gas} = P_{Gas}/((\gamma_{Gas} - 1)e_{Gas})$; the specific heat capacity $R_{Gas} = C_{P,Gas} - C_{V,Gas} = \gamma_{Gas}C_{V,Gas} - C_{V,Gas} = (\gamma_{Gas} - 1)C_{V,Gas}$, $e_{Gas} = C_{V,Gas}T_{Gas}$, thus $P_{Gas} = (\gamma_{Gas} - 1)\rho_{Gas}e_{Gas} = (\gamma_{Gas} - 1)\rho_{Gas}C_{V,Gas}T_{Gas} = \rho_{Gas}R_{Gas}T_{Gas}$, $\rho_{Gas} = P_{Gas}/(\gamma_{Gas} - 1)C_{V,Gas}T_{Gas} = P_{Gas}/(R_{Gas}T_{Gas})$.

Then the mixture density is

$$\rho = \alpha_{Gas1}\rho_{Gas1} + \alpha_{Gas2}\rho_{Gas2} + \alpha_L\rho_L = \alpha_{Gas1}\frac{P_{Gas1}}{(\gamma_{Gas1}-1)C_{V,Gas1}T_{Gas1}} + \alpha_{Gas2}\frac{P_{Gas2}}{(\gamma_{Gas2}-1)C_{V,Gas2}T_{Gas2}} + \alpha_L\frac{P_L+\gamma_L P_{\infty,L}}{(\gamma_L-1)(e_L-q_L)} \quad (A.1.2)$$

The internal energy for liquid is $e_L(P_L, \rho_L) = (P_L + \gamma_L P_{\infty,L})/((\gamma_L-1)\rho_L) + q_L$, internal energy for gas is $e_{Gas}(P_{Gas}, \rho_{Gas}) = P_{Gas}/(\gamma_{Gas}-1)\rho_{Gas}$, thus internal energy for fluid mixtures are

$$e = \xi_{Gas1}e_{Gas1} + \xi_{Gas2}e_{Gas2} + \xi_L e_L = \xi_{Gas1}C_{V,Gas1}T_{Gas1} + \xi_{Gas2}C_{V,Gas2}T_{Gas2} + \xi_L\left(\frac{P_L+\gamma_L P_{\infty,L}}{(\gamma_L-1)\rho_L} + q_L\right) \quad (A.1.3)$$

The speed of sound for liquid and gas could be obtained through $c_L = \sqrt{\gamma_L(P_L + P_{\infty,L})/\rho_L} = \sqrt{\gamma_L(P_L + P_{\infty,L})/(\rho\xi_L/\alpha_L)}$ and $c_{Gas} = \sqrt{\gamma_{Gas}P_{Gas}/\rho_{Gas}} = \sqrt{\gamma_{Gas}R_{Gas}T_{Gas}}$.

The volume fraction for components are

$$\alpha_{Gasi} = \xi_{Gasi}\frac{\rho}{\rho_{Gasi}} = \xi_{Gasi}\frac{\rho}{\frac{P_{Gasi}}{(\gamma_{Gasi}-1)e_{Gasi}}} = \xi_{Gasi}\frac{\rho}{\frac{P_{Gasi}}{(\gamma_{Gasi}-1)C_{V,Gasi}T_{Gasi}}} = \frac{\xi_{Gasi}\rho(\gamma_{Gasi}-1)C_{V,Gasi}T_{Gasi}}{P_{Gasi}} \quad (A.1.4)$$

$$\alpha_L = 1 - \sum \alpha_{Gasi} \quad (A.1.5)$$

### A.2.1. Peng-Robinson Equation of State
Besides, the "Peng-Robinson" EOS[118] is also combined into current four equation scheme and the related test case, near-critical shock droplet interaction, is presented in A.2.2.
For the liquid and gas components,

$$p = \frac{RT}{v-b} - \frac{a}{v^2+2bv-b^2} \quad (A.2.1)$$

where T is the temperature; R is the universal gas constant; V is the molar volume, V = M/ρ, M is the molar mass. Coefficients $a = \sum_{\alpha=1}^{N}\sum_{\beta=1}^{N} X_\alpha X_\beta a_{\alpha\beta}$ and $b = \sum_{\alpha=1}^{N} X_\alpha b_\alpha$. $X_\alpha$ is the mole fraction of species α and in-total species number is N; coefficients $a_{\alpha\beta} = 0.457236(RT_{c,\alpha\beta})^2/p_{c,\alpha\beta}(1 + c_{\alpha\beta}(1 - \sqrt{T/T_{c,\alpha\beta}}))^2$ and $b_\alpha = 0.077796RT_{c,\alpha}/p_{c,\alpha}$ are obtained according to the mixing rules[119]. $p_{c,\alpha\beta}$ is the critical mixture pressure and $p_{c,\alpha\beta} = Z_{c,\alpha\beta}RT_{c,\alpha\beta}/v_{c,\alpha\beta}$, $c_{\alpha\beta} = 0.37464 + 1.5422\omega_{\alpha\beta} - 0.26992\omega_{\alpha\beta}^2$, $T_{c,\alpha\beta}$ is the critical mixture temperature and $T_{c,\alpha\beta} = \sqrt{T_{c,\alpha}T_{c,\beta}}(1 - k_{\alpha\beta})$. $T_{c,\alpha}$ and $T_{c,\beta}$ are critical temperature for species α and β, $k_{\alpha\beta}$ is the binary interaction parameter. The critical mixture molar volume $v_{c,\alpha\beta}$, the critical mixture compressibility $Z_{c,\alpha\beta}$, the acentric factor $\omega_{\alpha\beta}$ denote as $v_{c,\alpha\beta} = \frac{1}{8}(v_{c,\alpha}^{1/3} + v_{c,\beta}^{1/3})^3$, $Z_{c,\alpha\beta} = \frac{1}{2}(Z_{c,\alpha} + Z_{c,\beta})$, $\omega_{\alpha\beta} = \frac{1}{2}(\omega_\alpha + \omega_\beta)$, where $v_{c,\alpha}^{1/3}$ and $v_{c,\beta}^{1/3}$ are critical molar volume for species α and β, $Z_{c,\alpha}$ and $Z_{c,\beta}$ are critical compressibility factor for species α and β, $\omega_\alpha$ and $\omega_\beta$ are acentric factor for species α and β.
Besides, reference[120] provides the parameters for the NASA polynomials, which would be used to obtain the internal energy, enthalpy, and entropy.

### A.2.2. PR-EOS for (near-critical) Shock Droplet Interaction
We validate the shock interaction in a nitrogen environment with a sphere of n-Dodecane droplet[121] in order to show the performance of current numerical scheme in complex realistic conditions. Same parameters are adopted as the reference except for the mesh resolution. The uniform mesh size(0.115mm) is adopted near the droplet, which is finer than that of reference(0.23mm). In this way, it should be noted that higher mesh resolution is adopted for current scheme in order to be comparable to results from high order WENO5 scheme of reference. Ref.[120] provides the parameters for the NASA polynomials. Numerical scheme could be found in the Appendix B.
In Fig.A.2.1, results show good agreement with those from reference[121]. It shows that current scheme could be combined with real fluid model directly, further proving its robust performance. Moreover, based on same meth resolution(0.115mm) and same initial conditions, current results are superior to these of WENO3(HLLC) and OWENO3(HLLC) scheme, especially near the two phase interface, which simply shows its comparable performance over higher order method.

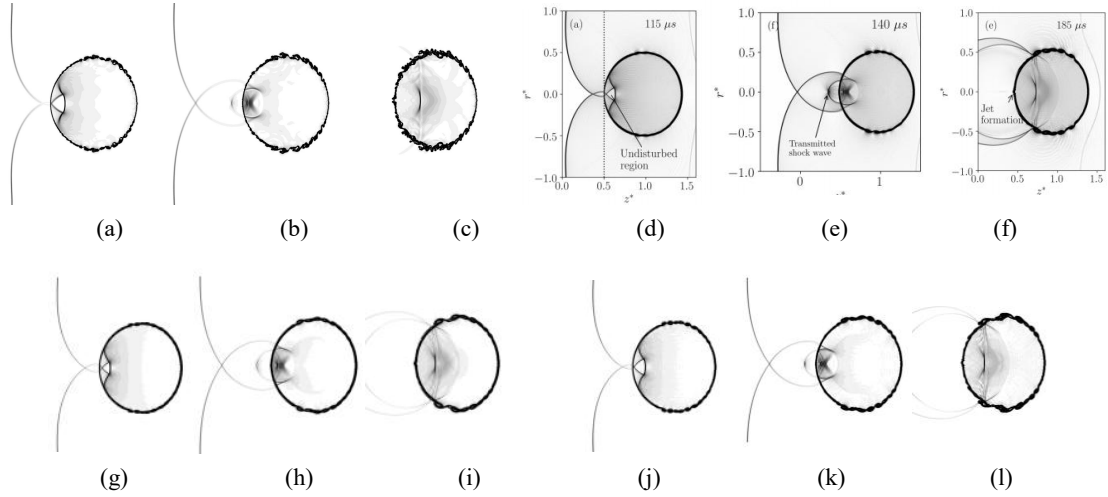

Figure.A.2.1.Shock n-Dodecane in the Nitrogen environment:115μs,(a)(d)(g)(j);140μs,(b)(e)(h)(k);185μs,(c)(f)(i)(l).

Current results(a~c);Reference(d~f)[121];WENO3(g,h,i);OWENO3(j,k,l).

## Appendix B. Internal energy and real fluid effects

Internal energy, as an additional parameter, needs to be taken into account. Most procedures, such as step adopting All-Mach Riemann solver and reconstruction steps for mass fraction, are the same as those in Chapter3. While following aspects are excepted.

The thermodynamic model adopted here is related to Appendix.A. Variables including velocities, pressure, mass fraction and temperature are given, others like density and internal energy could be obtained according to the thermodynamic relations, $\rho = f(p, T, \xi_{Gasi})$, $e = f(\rho, T, \xi_{Gasi})$.

In a general way, variables $u^*, v^*, w^*, \rho^*, p^*, \xi^*_{Gasi}$ adopt the same reconstruction method as explained in Chapter3 and internal energy $e^*$ just need follow the reconstruction method of density $\rho^*$. While in the current scheme, after obtaining the interface internal energy $e^*$, the TDU idea would be applied to update variables like density $\rho^*$. Specifically, during the reconstruction process, interface parameters like density are updated from the interface pressure, as mass fraction and internal energy, $\rho^* = f(p^*, \xi^*_{Gasi}, e^*)$, thus the numerical format of density keeps numerical consistent with other variables. Moreover, volume fraction could be obtained according to $\alpha^*_i = f(\rho^*, p^*, \xi^*_{Gasi}, e^*)$.

If the current FC scheme is combined with thermodynamics relations from Appendix.A.2.1, temperature is updated firstly according to internal energy, density, species mass fraction $T = f(e, \rho, \xi_{Gasi})$ (by Gradient descent/Newton method), then pressure is updated from temperature, density and species mass fraction $p = f(\rho, T, \xi_{Gasi})$. A small note is that the convergence speed of the Gradient descent/Newton method in obtaining the temperature could be improved by storing the temperature from the previous time step and using it as the initial guess to calculate the updated temperature. In the FC scheme, the numerical flux evaluated from the two neighboring cells of the face is exactly the same, which guaranty strict conservation of all variables.

## Appendix C. A spherical drop in static equilibrium

A spherical drop in static equilibrium is adopted to show the balanced effects between pressure and surface tension force since imbalance between them induce the parasitic velocities. Parameters are adopted according to the classical reference[122]. Density ratio of two in-viscid fluids is 10. The side length of computational cube domain is 8 and 40 uniform meshes are applied in every direction. Initial pressure drop between inside fluid and outside fluid is equal to the theoretical equilibrium value, $\sigma = 73, R = 2$, thus $p_{inside} - p_{outside} = 2\sigma/R = 73$.

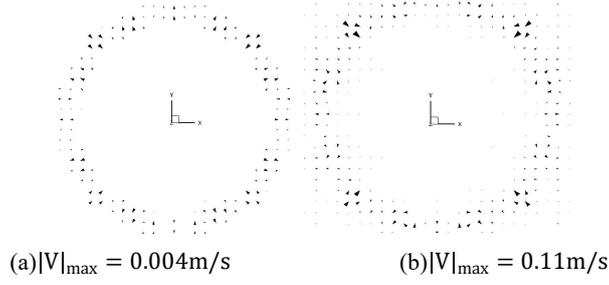

(a)$|V|_{max} = 0.004 m/s$  (b)$|V|_{max} = 0.11 m/s$

Figure C.1: The velocity vector (vector length is grid units divide magnitude) field of an x-y plane. Plots (a) is at t≈0.001s with vector length 40, (b) is at t≈0.05s with vector length 1.5.

For our current method, the max-mum parasitic speed at t≈0.001s and t≈0.05s are about 0.004m/s and 0.11m/s respectively, which is comparable to best results 0.0855m/s and 0.386m/s shown in the reference [122]. Thus, current surface tension model performs well.

## Appendix D. High order improvement with turbulence modeling(iLES)

The thermodynamic model applied for this part is related to Section2.3. In order to include the turbulence modeling, a high order improvement of All-Mach THINC-TDU is proposed with iLES approach, two extra gas components as well as modified sensor. For the iLES scheme of the reference[23], if discontinues region is detected by sensor, upwind-basied reconstruction (density) would be adopted. While in current scheme, the cell interface value reconstruction procedure is switched among an upwind-biased, a centered reconstruction and a THINC-based scheme. And TDU idea is adopted to continually keep the thermodynamic relationship coupled among variables(pressure, density, gas mass fraction).

- **Upwind-biased reconstruction scheme**

Procedures(Regions without discontinuous interfaces) shown in 3.3 are adopted for the upwind-biased reconstructions of variables.

- **Central reconstruction scheme**

The above upwind-biased scheme has intrinsic numerical dissipation, thus the central reconstruction scheme is needed to reduce dissipation. And proper sensor function is used to switch between candidate schemes. A linear fourth order central scheme is applied,

$$\Phi^{*,C}_{i+1/2} = [u^*, v^*, w^*, p^*, \xi^*_{Gas1}, \xi^*_{Gas2}]^C_{i+1/2} = 1/12[7(\overline{\Phi_i} + \overline{\Phi_{i+1}}) - \overline{\Phi_{i-1}} - \overline{\Phi_{i+2}}] \quad (D.1)$$

$$\Phi^{*,C}_{i+1/2} = [\rho^*]^C_{i+1/2} = 1/2[(\Phi^*_{I-1/2} + \Phi^*_{I+1/2})] \quad (D.2)$$

- **THINC-based reconstruction scheme**

In order to compute the flux across the cell interface between cells 2 and 3, we reconstruct left and right hand side values of the scalar $\xi_{Gas}$ as the equations from **(3-3) to (3-4)** shown in **3.3**. It's also noted that discontinuous two-phase interface region generally generally meet the requirements $\varepsilon < \overline{\xi_i} < 1 - \varepsilon$ and $(\overline{\xi_i} - \overline{\xi_{i-1}})(\overline{\xi_{i+1}} - \overline{\xi_i}) > 0$, where ε could be a small positive parameter comparing with $\overline{\xi_i}$.

- **Modified sensor**

This newly modified discontinuity sensor is used to identify the discontinues region. The flow field smoothness is identified via a function of smooth sensor, which works as a switch to change the type of the reconstruction method for the primitive variables u, v, w, ρ, p, $\xi_{Gas1}, \xi_{Gas2}$ on the cell interface. The primitive variables on the cell interface denote:

$$\Phi^*_{i+1/2} = [1 - f(\theta)]\Phi^{smooth} + f(\theta)\Phi^{discontinues} = [1 - f(\theta)]\Phi^{*,C}_{i+1/2} + f(\theta)((1 - \sigma)\Phi^{*,U}_{i+1/2} + \sigma\Phi^{*,Th}_{i+1/2}) \quad (D.3)$$

σ is a Dirichlet function, which could be adjustable according to needed: generally for reconstruction step of pressure and velocity it becomes zero; for reconstruction step of two phase interface region, it becomes one while for other region it becomes zero.

Specifically, we use vorticity-dilation sensor developed by Ducros et al.[123] to detect the compressible shock and expansion waves in present fully compressible flow: $\theta^D = (\nabla \cdot u)^2/((\nabla \cdot u)^2 + (\nabla \times u)^2 + \epsilon)$, where, $\epsilon$ denotes a very small value ($\epsilon = 10^{-20}$), making denominator non-zero.

Two-phase interface are detected with the variation of the total gas volume fraction in all three spatial directions: $\theta^{Gas} = var_i(Gas) + var_j(Gas) + var_k(Gas)$ , where, $var_i(Gas) = \|\alpha_{Gas1, i} + \alpha_{Gas2, i} - (\alpha_{Gas1, i-1} + \alpha_{Gas2, i-1})\| + \|\alpha_{Gas1, i+1} + \alpha_{Gas2, i+1} - (\alpha_{Gas1, i} + \alpha_{Gas2, i})\|$. Same expressions are adopted for vapor volume fraction and its detector is denoted as $\theta^\alpha$.

Subsequently, we give the switch criteria:

$$f(\theta^D, \theta^\alpha, \theta^{Gas}) = \begin{cases} 1, if\ \theta^D > \theta_{th}^D\ ||\ \theta^\alpha > \theta_{th}^\alpha\ ||\ \theta^{Gas} > \theta_{th}^{Gas} \\ 0,\ others \end{cases} \quad \text{(D.4)}$$

It means that if one or more sensor exceeds its threshold value ($\theta^D > \theta_{th}^D$ or/and $\theta^\alpha > \theta_{th}^\alpha$ or/and $\theta^{Gas} > \theta_{th}^{Gas}$), the numerical scheme switches to the discontinues reconstructions $\Phi_{i+1/2}^* = \Phi^{discontinues} = (1-\sigma)\Phi_{i+1/2}^{*,U} + \sigma\Phi_{i+1/2}^{*,Th}$. The threshold values $\theta_{th}^D = 0.95$ is suggested by Egerer et al.[23] and $\theta_{th}^{Gas} = 0.4$ is for the volume fraction of total gas phase suggested by Trummler et al. [22]. And these proposed modifications could help to avoid pressure oscillations while keeping contact waves crisp without artificial smearing.

Specifically, for smooth region, velocity and pressure adopt the linear 4th order reconstruction process, while a linear 2nd order central approximation reconstruction is implemented for the density, internal energy(if have), as well as two additional mass fraction field. For discontinues fields, a upwind biased reconstruction is generally implemented $\Phi^{discontinues} = \Phi_{i+1/2}^{*,U}$. Generally, velocity components are reconstructed using the third order slope limiter Koren [100] and the thermodynamic quantities density, pressure, internal energy(if have), two additional mass fraction are reconstructed using the second order Minmod slope limiter (Roe,1986)[99]. It should be noted that the new mass fraction $\xi$ generally keep the same reconstruction scheme of density, $\xi^{discontinues} = \xi_{i+1/2}^{*,U}$. While in the discontinuous region of two phase interface, instead of using upwind biased reconstruction with slop limiter for two additional mass fraction(or density), THINC is combined for mass fraction(or density) $\xi^{discontinues} = (1-\sigma)\xi_{i+1/2}^{*,U} + \sigma\xi_{i+1/2}^{*,Th}$, and TDU idea is then adopted for density(or mass fraction) $\rho^{discontinues} = f(\xi^{discontinues}, p^{discontinues})$. In this way, variables are updated consistently, interface is sharpened and pressure oscillation is avoided.

Besides, the transport velocity and interface pressure are represented as

$$u^* = (1-f(\theta))(u_{i+1/2}^{*,C} - \frac{\Delta^3 p_{i+1/2}^*}{I_L+I_R}) + f(\theta)\frac{I_L u_L + I_R u_R + p_L - p_R}{I_L+I_R} \quad \text{(D.5)}$$

$$p^* = (1-f(\theta))p_{i+1/2}^{*,C} + f(\theta)(\frac{p_L+p_R}{2}) \quad \text{(D.6)}$$

$I_L = ((3\overline{\rho_i} + \overline{\rho_{i+1}}) * max(c_{l,i}, c_{l,i+1}))/4$, $I_R = ((\overline{\rho_i} + 3\overline{\rho_{i+1}}) * max(c_{l,i}, c_{l,i+1}))/4$, $\Delta^3 p_{i+1/2}^*$ is an approximation of the third pressure derivative, the acoustic impedances are $I_L$ and $I_R$, $max(c_{l,i}, c_{l,i+1})$ is the maximum liquid speed of sound.

Other procedures (Remarks) are consistent with Charpter3. Through complete implementation, liquid-liquid vapor and liquid-gas two phase interface are sharpened and dispassion of discontinues region are reduced.

## Acknowledgement

This research is supported by EDEM project and EDEM project has received funding from the European Union Horizon 2020 Research and Innovation programme. Grant Agreement No 861002. The authors also gratefully acknowledge the Leibniz Supercomputing Centre for funding this research by providing computing time on its Linux-Cluster.